\documentclass[letterpaper,11pt]{article}
\pdfoutput=1

\usepackage{hyperref}
\hypersetup{
    colorlinks=true,       % false: boxed links; true: colored links
    linkcolor=blue,          % color of internal links
    citecolor=blue,        % color of links to bibliography
    filecolor=blue,      % color of file links
    urlcolor=blue           % color of external links
}

\usepackage{float}
                     
\usepackage{braket,slashed,bm}
\usepackage{array,multirow}
\usepackage[normalem]{ulem}
\usepackage{xcolor,cancel,youngtab}

\usepackage[T1]{fontenc} % if needed

\usepackage{mathrsfs}
\usepackage{booktabs}
\usepackage{adjustbox}
\usepackage{mathtools}

\usepackage{soul}

\usepackage{tikz}
\usetikzlibrary{arrows,decorations.pathmorphing,backgrounds,positioning,fit,petri,automata,shadows,calendar,mindmap,
decorations.markings,calc}

\definecolor{labelkey}{rgb}{0,0.5,0.0}

\usepackage{xspace}
\usepackage{slashed}
\usepackage{array,multirow}
\usepackage{graphicx}
\usepackage{graphics}
\usepackage{epsfig}
\usepackage{amsfonts}
\usepackage{amsmath}
\usepackage{amssymb}
\usepackage{dsfont}
\usepackage{color}
\usepackage{xcolor}
\usepackage{cite}
\usepackage{pdflscape}
\usepackage{pifont}
\usepackage{pgfplots}
\usepackage{tikz} 
\usepackage{pgfplotstable}
\usepackage{bbold}

\newcommand{\hc}{\mathrm{h.c.}}
\newcommand{\ep}{\epsilon}

\newcommand{\al}{\alpha}
\newcommand{\bt}{\beta}

\newcommand{\dt}{\delta}

\newcolumntype{P}[1]{>{\centering\arraybackslash}p{#1}}

\usepackage{bm}  %                                          

\newcommand{\beq}{\begin{equation}}
\newcommand{\eeq}{\end{equation}}
\newcommand{\be}{\begin{equation}}
\newcommand{\ee}{\end{equation}}
\newcommand{\bea}{\begin{eqnarray}}
\newcommand{\eea}{\end{eqnarray}}
\newcommand{\ben}{\begin{eqnarray*}}
\newcommand{\een}{\end{eqnarray*}}

\renewcommand{\vec}[1]{{\mathbf #1}} 

\newcommand{\bma}{\begin{pmatrix}}
\newcommand{\ema}{\end{pmatrix}}

\def\lixo#1{}

\def\slashchar#1{\setbox0=\hbox{$#1$}           % set a box for#1
  \dimen0=\wd0                                    % and get its size
  \setbox1=\hbox{/} \dimen1=\wd1                  % get size of/
  \ifdim\dimen0>\dimen1                           % #1 is bigger
    \rlap{\hbox to \dimen0{\hfil/\hfil}}            % so center / in box
    #1                                             % and print #1
  \else                                          % / is bigger
    \rlap{\hbox to \dimen1{\hfil$#1$\hfil}}        % so center #1
    /                                           % and print/
 \fi}                                           %

\newcommand{\Or}{\mathcal O}

\newcommand{\sq}{^{2}}

\newcommand{\dslash}[1]{#1 \llap{/\kern-0.5pt}}
\newcommand{\Dslash}[1]{#1 \llap{/\kern+1.5pt}}
\newcommand{\DDslash}[1]{#1 \llap{/\kern+2.3pt}}
\newcommand{\dslashh}[1]{#1 \llap{/\kern+1pt}}

\newcommand{\NLDBD}{$0 \nu \beta \beta$}

\definecolor{cadmiumgreen}{rgb}{0.0, 0.42, 0.24}
\definecolor{darkpastelgreen}{rgb}{0.01, 0.75, 0.24}
\definecolor{darkspringgreen}{rgb}{0.09, 0.45, 0.27}
\definecolor{forestgreen(web)}{rgb}{0.13, 0.55, 0.13}
\definecolor{forestgreen(traditional)}{rgb}{0.0, 0.27, 0.13}
\definecolor{cobalt}{rgb}{0.0, 0.28, 0.67}
\definecolor{darkblue}{rgb}{0.0, 0.0, 0.75}
\definecolor{darkred}{rgb}{0.55, 0.0, 0.0}
\definecolor{palatinatepurple}{rgb}{0.41, 0.16, 0.38}
\definecolor{burntorange}{rgb}{0.8, 0.33, 0.0}

\newcommand{\nn}{\nonumber}
\newcommand{\nnl}{\nonumber \\}

\newcommand{\cL}{{\mathcal L}}

\newcommand{\eS}{\epsilon_S}
\newcommand{\eT}{\epsilon_T}
\newcommand{\eP}{\epsilon_P}
\newcommand{\eL}{\epsilon_L}
\newcommand{\eR}{\epsilon_R}

\newcommand{\teS}{{\tilde{\epsilon}_S}}
\newcommand{\teT}{{\tilde{\epsilon}_T}}
\newcommand{\teP}{{\tilde{\epsilon}_P}}
\newcommand{\teL}{{\tilde{\epsilon}_L}}
\newcommand{\teR}{{\tilde{\epsilon}_R}}

\def\12{\frac{1}{2}}
\def\0nbb{$0\nu\beta\beta$}

\begin{document}
	
\begin{titlepage}

\begin{flushright}
%	\\
\end{flushright}

\vspace{2.0cm}

\begin{center}
	{\LARGE  \bf 
Sterile neutrinos with non-standard interactions \\
		\vspace{3mm}
		in $\beta$- and $0\nu\beta\beta$-decay experiments %\\ 
	}
	\vspace{1cm}
	
	{\large \bf  W. Dekens$^{a}$, J. de Vries$^{b,c}$, T. Tong$^{b,d}$} 
	\vspace{0.5cm}

	\vspace{0.25cm}
	
	\vspace{0.25cm}
	{\large 
		$^a$ 
		{\it 
			Department of Physics, University of California at San Diego, La Jolla, CA 92093, USA}}
	
{\large 
$^b$ 
{\it Institute for Theoretical Physics Amsterdam and Delta Institute for Theoretical Physics, University of Amsterdam, Science Park 904, 1098 XH Amsterdam, The Netherlands}}

{\large 
$^c$ 
{\it Nikhef, Theory Group, Science Park 105, 1098 XG, Amsterdam, The Netherlands}}

	\vspace{0.25cm}
	{\large 
		$^d$ 
		{\it 
		Institute for Fundamental Science and Department of Physics, University of Oregon, Eugene, OR 97403}}
	
\end{center}

\vspace{0.2cm}

\begin{abstract}
\vspace{0.1cm}
Charged currents are probed in low-energy precision $\beta$-decay experiments and at high-energy colliders, both of which aim to measure or constrain signals of beyond-the-Standard-Model physics. In light of future $\beta$-decay and LHC measurements that will further explore these non-standard interactions, we investigate what neutrinoless double-$\beta$ decay ($0\nu\beta\beta$) experiments can tell us if a nonzero signal were to be found. Using a recently developed effective-field-theory framework, we consider the effects that interactions with right-handed neutrinos have on $0\nu\beta\beta$ and
 discuss the range of neutrino masses that current and future $0\nu\beta\beta$ measurements can probe, assuming neutrinos are Majorana particles. 
For non-standard interactions at the level suggested by recently observed hints in $\bt$ decays, we show that next-generation $0\nu\beta\beta$ experiments can determine the Dirac or Majorana nature of neutrinos, for sterile neutrino masses larger than $\mathcal O(10)$ eV.

\end{abstract}

\vfill

\end{titlepage}

\tableofcontents

\section{Introduction}
Investigations of $\beta$-decay processes have played an important role in the development of the Standard Model (SM) of particle physics. In the modern era, precision analyses of pion, neutron, and nuclear $\beta$-decay are used to extract precise values of SM parameters (in particular $V_{ud} G_F$) and to probe beyond-the-Standard-Model (BSM) physics. 
At the relatively low energies associated with $\beta$-decay experiments, which involve $Q$-values around an MeV, BSM physics can be described using effective-field-theory (EFT) techniques as was already proposed by Lee and Yang \cite{LeeYang}, for recent reviews see e.g.\ Refs.~\cite{Gonzalez-Alonso:2013uqa,Cirigliano:2013xha,Vos:2015eba,Gonzalez-Alonso:2018omy,Bischer:2019ttk,Hardy:2020qwl}. Within this framework the $V-A$ structure of the SM charged weak interaction is described by a dimension-six operator, while deviations are captured by additional higher-dimensional operators which are suppressed by powers of $v/\Lambda$, where $v\simeq 246$ GeV is the electroweak scale  and $\Lambda$ the scale of new physics. At the leading dimension-six level, the SM current is then augmented by ten effective interactions each of which comes with an, in principle complex, Wilson coefficient that scales as $v^2/\Lambda^2$. 

These charged-current interactions can be investigated both at high energies, e.g.\ in $pp\to e\nu$ at the LHC, as well as in low-energy precision measurements.
Focussing on probes in the latter category, Falkowski, Gonz\'alez-Alonso, and Naviliat-Cuncic recently performed a state-of-the-art analysis of neutron and nuclear $\beta$-decay data \cite{Falkowski:2020pma} (we will refer to this work by FGN). In addition to neutron decay and superallowed $0^+\rightarrow 0^+$ transitions, FGN included data from mirror $1/2^+ \rightarrow 1/2^+$ decays. The authors then performed a simultaneous fit including all leading-order dimension-six Wilson coefficients (considering only the real parts), while taking into account theory uncertainties by marginalizing over nuisance parameters related to radiative corrections and nuclear matrix elements (NMEs).
Armed with this framework, FGN then analyzed all relevant data in several scenarios. Assuming only SM interactions FGN obtained accurate values for $V_{ud} = 0.97370(25)$ and $g_A = 1.27276(45)$, illustrating that $\beta$-decay experiments can be used to extract precise determinations of fundamental SM parameters and hadronic matrix elements. Adding non-standard interactions involving only left-handed neutrinos to the global fit gives an impressive confirmation of the SM: non-standard scalar and tensor interactions are constrained at the per-mille level, corresponding to scales of BSM physics of several TeV (note that pseudoscalar interactions give suppressed contributions and were not considered). 
Assuming the BSM scale is well above the electroweak scale, LHC measurements can be used to derive similar, complementary, constraints \cite{Gonzalez-Alonso:2013uqa,Cirigliano:2013xha}.

This picture changes somewhat once non-standard interactions involving right-handed neutrinos are included as well. In this case, the global fit constrains vector, axial-vector, and scalar couplings to right-handed neutrinos at the few percent level, but prefers a nonzero right-handed tensor coupling at the $10\%$ level (about 3.2$\sigma$ away from the SM point). This discrepancy is driven by a single recent measurement of the ``little-a'' coefficient in neutron decay \cite{aspect} and is not significant enough to warrant too much excitement. In particular it is unclear whether one can construct a viable BSM scenario, given the current LHC constraints from $pp\to e\nu$ \cite{Gonzalez-Alonso:2013uqa,Cirigliano:2013xha}.
Nevertheless, in light of the FGN result, the recent hints for BSM effects in CKM unitarity tests  \cite{Grossman:2019bzp,Crivellin:2020ebi,Crivellin:2020lzu,Crivellin:2021njn}, and more generally in context of ongoing and future $\beta$-decay experiments and collider measurements,  it is interesting to ask what a signal of BSM charged-currents would imply for other complementary experiments. For instance, Refs.~\cite{Gonzalez-Alonso:2013uqa,Cirigliano:2013xha} demonstrated a strong complementarity between $\beta$-decay and LHC observables.

Here we focus on couplings to right-handed neutrinos and the connection to probes of lepton number violation (LNV). The presence of right-handed neutrinos in nature would imply the existence of fields that are neutral under the full SM gauge group~\footnote{They do not have be singlets under BSM gauge groups. For example, in left-right symmetric models \cite{Pati:1974yy, Mohapatra:1974hk, Senjanovic:1975rk} the right-handed neutrinos are charged under $SU(2)_R$ and couple to right-handed gauge bosons. At low energies, the right-handed neutrinos are sterile with respect to the SM gauge group, but interact through higher-dimension couplings that arise from integrating out the heavy BSM fields.}. As a result, nothing forbids the existence of a Majorana mass for fields like this unless additional symmetries are invoked. Such a mass term would imply that the mass eigenstates of neutrinos are Majorana and the violation of Lepton Number ($L$) by two units. So, if $\beta$-decay or collider experiments uncover evidence for right-handed neutrino interactions, a reasonable follow-up question would be: in which scenarios can we expect to find correlated signals in experimental probes of LNV? Here, we investigate this question in the context of searches for neutrinoless double beta decay ($0\nu\beta\beta$), the process where two neutrons in a nucleus transform into two protons with the emission of two electrons but zero (anti-)neutrinos. 

To compare constraints from $\beta$- and $0\nu\beta\beta$-decay experiments on couplings to right-handed neutrinos we need to consider the mass scale of right-handed neutrinos. $\beta$ decays are insensitive to neutrino masses whenever the neutrinos are light, $m_i\ll Q$, while a $0\nu\beta\beta$ signal is proportional to the Majorana masses of the neutrinos. We therefore focus on right-handed neutrinos with masses well below the MeV scale that can be considered approximately massless for $\beta$ decay experiments (as well as at colliders).
Additional constraints on sterile neutrinos could in principle come from cosmological  and astrophysical probes \cite{Adhikari:2016bei,Bolton:2019pcu} such as big bang nucleosynthesis \cite{Boyarsky:2009ix,Ruchayskiy:2012si} considerations or the cosmic microwave background  \cite{Vincent:2014rja}. Such constraints however depend on the thermal history of the universe and can be avoided in specific scenarios \cite{Nemevsek:2012cd}.

In what follows, we calculate the expected $0\nu\beta\beta$ decay rates of various isotopes as a function of the non-standard couplings and the right-handed neutrino mass. Barring significant cancelations between the contributions from the `standard mechanism' and those from sterile neutrinos, we then discuss the range of neutrino masses for which a $\beta$-decay or collider signal could imply a measurable signal in $0\nu\beta\beta$.
We start by setting up the EFT framework in Sect.~\ref{nuSMEFT} and  briefly describe the FGN analysis in  Sect.~\ref{FGN}. The $0\nu\beta\beta$ decay rates in the presence of light right-handed neutrinos with non-standard interactions are discussed in Sect.~\ref{0vbb}. We present our analysis of EFT couplings in Sect.~\ref{results} and consider a specific BSM model involving leptoquarks in Sect.~\ref{LQ}. We conclude in Sect.~\ref{conclusion}. Appendices are devoted to the matching relations with the neutrino-extended Standard Model Effective Field Theory and the calculation of $0\nu\beta\beta$ decay rates.

\section{The Lagrangian in the Standard Model Effective Field Theory}\label{nuSMEFT}
In this work, we focus on interactions that give rise to $\beta$ decays, especially those that involve right-handed neutrinos. We therefore consider an effective field theory that consists of the SM supplemented by an SM gauge singlet, $\nu_R$, and include higher-dimensional operators up to dimension six. The Lagrangian of the resulting EFT, called the neutrino-extended Standard Model EFT ($\nu$SMEFT) can then be written as
\begin{eqnarray}\label{eq:smeft}
	\mathcal L &=&  \mathcal L_{SM} + \bar \nu_R\, i\slashed{\partial}\nu_R- \left[ \frac{1}{2} \bar \nu^c_{R} \,\bar M_R \nu_{R} +\bar L \tilde H Y_\nu \nu_R + \rm{h.c.}\right]\nn \\
	&&+  \mathcal L^{(\bar 5)}_{\nu_L}+  \mathcal L^{(\bar 5)}_{\nu_R}+  \mathcal L^{(\bar 6)}_{\nu_L}+  \mathcal L^{(\bar 6)}_{\nu_R} \,,
\end{eqnarray}
where we used $\Psi^c = C \bar \Psi^T$ for a field $\Psi$ in terms of the charge conjugation matrix $C = - C^{-1} = -C^T = - C^\dagger$. We use the definition for chiral fields $\Psi_{L,R}^c = (\Psi_{L,R})^c =  C \overline{\Psi_{L,R}}^T= P_{R,L} \Psi^c$, with $P_{R,L}=(1\pm\gamma_5)/2$.
Furthermore,  $L=(\nu_L,\, e_L)^T$ is the left-handed lepton doublet and $\tilde H = i \tau_2 H^*$ with $H$ the Higgs doublet 
\begin{equation}
	H = \frac{v}{\sqrt{2}} U(x) \left(\begin{array}{c}
		0 \\
		1 + \frac{h(x)}{v}
	\end{array} \right)\,.
\end{equation}
Here $v=246$ GeV is the Higgs vacuum expectation value (vev),  $h(x)$ is the Higgs field, and $U(x)$ is an $SU(2)$ matrix encoding the Goldstone modes. Generally, $\nu_{R}$ is a column vector of $n$ right-handed sterile neutrinos, making $Y_\nu$ a $3\times n$ matrix of Yukawa couplings and $\bar M_R$ a symmetric complex $n \times n$ mass matrix. 
We will work in  the basis where the charged leptons $e^i_{L,R}$ and quarks $u^i_{L,R}$ and $d^i_R$ are mass eigenstates ($i=1,2,3$). After electroweak symmetry breaking this then gives $d^i_L = V^{ij} d_L^{j,\,\rm mass}$, where $V$ is the CKM matrix. We discuss the mass and flavor bases for the neutrinos in the next subsection. 

Finally, the relevant dimension-five operators can be written as,
\be
\mathcal L^{(\bar 5)}_{\nu_L} = \ep_{kl}\ep_{mn}(L_k^T\, C^{( 5)}\,CL_m )H_l H_n\,,\qquad  \mathcal L^{(\bar 5)}_{\nu_R}=- \bar \nu^c_{R} \,\bar M_R^{(5)} \nu_{R} H^\dagger H\,,
\ee
while the needed dimension-six operators involving left- and right-handed neutrinos are collected in Tables \ref{tab:O6L} and \ref{tab:O6R} of appendix \ref{app:match}, respectively.

\subsection{The Lagrangian below the electroweak scale}\label{massbasis}
Below the electroweak scale the Lagrangian of Eq.\ \eqref{eq:smeft} induces the following mass terms for the neutrinos,
  \bea\label{eq:numasses}
 \mathcal L_m = -\frac{1}{2} \bar N^c M_\nu N +{\rm h.c.}\,,\qquad M_\nu = \bma M_L &M_D^*\\M_D^\dagger&M_R^\dagger \ema \,,
 \eea
where $M_L=-v^2 C^{(5)}$, $M_D = \frac{v}{\sqrt{2}} \left[Y_\nu-\frac{v^2}{2} C_{L\nu H}^{(6)}\right]$,  and $M_R = \bar M_R+v^2 \bar M_R^{(5)}$. Furthermore, $N = (\nu_L,\, \nu_R^c)^T$ and $M_\nu$ is a $N\times N$ symmetric matrix, with $N=3+n$. This matrix can be diagonalized by an $N\times N$ unitary matrix, $U$, 
\bea\label{Mdiag}
U^T M_\nu U =m_\nu = {\rm diag}(m_1,\dots , m_{3+n})\,, \qquad N = U N_m\,.
\eea
This allows us to write the kinetic and mass terms of the neutrinos in the simple form
\bea
\mathcal L_\nu = \frac{1}{2} \bar \nu i\slashed \partial \nu -\frac{1}{2} \bar \nu^{ } m_\nu \nu\,,
\eea
in terms of the Majorana mass eigenstates $\nu = N_m +N_m^c = \nu^c$. These eigenstates are related to the flavor basis by
\bea\label{eq:numassbasis}
\nu_L = P_L(P U) \nu \,,\qquad \nu_L^c =P_R (P U^*) \nu\,,\nn\\
\nu_R =P_R (P_s U^*) \nu \,,\qquad \nu_R^c = P_L(P_s U) \nu\,,
\eea
where $P$ and $P_s$ are $3\times N$ and $n \times N$ projector matrices
\be
P = \begin{pmatrix}\mathcal I_{3\times 3} & 0_{3 \times n}  \end{pmatrix}\,,\qquad
P_s = \begin{pmatrix} 0_{n\times 3} & \mathcal I_{n \times n}  \end{pmatrix}\, ,
\ee
which project onto the active and sterile states, respectively.

We can use the above to write the charged-current interactions, which are induced by the dimension-six operators in Eq.\ \eqref{eq:smeft} below the electroweak scale, in the mass basis of the neutrinos. In the notation of Ref.\ \cite{Dekens:2020ttz}, this gives, at the scale $\mu \simeq 2$~GeV,
\bea\label{6final}
\mathcal L^{(6)}& =& \frac{2 G_F}{\sqrt{2}} \Bigg\{ 
  \bar u_L \gamma^\mu d_L \left[  \bar e_{R}  \gamma_\mu C^{(6)}_{\textrm{VLR}} \,  \nu+ \bar e_{L}  \gamma_\mu  C^{(6)}_{\textrm{VLL}} \,  \nu \right]+
  \bar u_R \gamma^\mu d_R \left[\bar e_{R}\,  \gamma_\mu  C^{(6)}_{\textrm{VRR}} \,\nu+\bar e_{L}\,  \gamma_\mu   C^{(6)}_{\textrm{VRL}} \,\nu  \right]\nn\\
& & +
  \bar u_L  d_R \left[ \bar e_{L}\, C^{(6)}_{ \textrm{SRR}}  \nu +\bar e_{R}\,  C^{(6)}_{ \textrm{SRL}}  \nu \right]+ 
  \bar u_R  d_L \left[ \bar e_{L} \, C^{(6)}_{ \textrm{SLR}}    \nu + \bar e_{R} \,  C^{(6)}_{ \textrm{SLL}}    \nu \right]\nn\\
&&+  \bar u_L \sigma^{\mu\nu} d_R\,  \bar e_{L}  \sigma_{\mu\nu} C^{(6)}_{ \textrm{TRR}} \, \nu+  \bar u_R \sigma^{\mu\nu} d_L\,  \bar e_{R}  \sigma_{\mu\nu}  C^{(6)}_{ \textrm{TLL}} \, \nu
\Bigg\}  +{\rm h.c.}\eea
The matching of the above coefficients to those in Eq.\ \eqref{eq:smeft} is described in App.\ \ref{app:match}. In the SM, only the $C_{\rm VLL}^{(6)}$ coefficient is nonzero. In the limit of vanishing neutrinos masses, the couplings $C_{\rm \Gamma A L}^{(6)}$, with $\Gamma =\{ V,S,T\}$ and $A=\{L,R\}$, are induced by the dimension-six interactions involving left-handed neutrinos in $\mathcal L^{(\bar 6)}_{\nu_L}$, while $C_{\rm \Gamma A R}^{(6)}$ are induced by the operators involving right-handed neutrinos in $\mathcal L^{(\bar 6)}_{\nu_R}$~\footnote{As can be seen from the matching in App.\ \ref{app:match}, this is no longer true for nonzero $M_{D,R}$ as these terms allow for chirality flips of the neutrino fields.}.

Eq.\ \eqref{6final} provides the interactions that are relevant for both single- and double-$\bt$ decays. As such, the above Lagrangian has been studied extensively in the literature in the context of $\bt$ decays, although usually using a different convention. We provide the translation to the notation of Ref.\ \cite{Falkowski:2020pma} in Table \ref{tab:translation}. Since $\bt$ decays are low-energy observables, they are commonly described in terms of the interactions involving nucleons instead of the quark-level interactions given above. The Lagrangian, often used in studies of $\bt$ decays, is then written in the following form~\footnote{The $C_\al^\pm$ couplings of Ref.\  \cite{Falkowski:2020pma} are defined in the flavor basis and thus related to those used here by the change of basis in Eq.\ \eqref{eq:numassbasis}. In our numerical analyses, we will often focus on the case with $n=1$ and $U_{4\al} = \dt_{4\al}$ for simplicity. In this scenario the $C_\al^-$ couplings used here coincide with those of FGN. }:
\bea
\label{eq:TH_Lleeyang}
\cL_{\rm Lee-Yang} &=& 
-  \bar{p}\gamma^\mu n 
\left(  \bar{e}_L \gamma_\mu C_V^+ \nu
+  \bar{e}_R \gamma_\mu C_V^-\nu \right) 
-  \bar{p}\gamma^\mu \gamma_5 n 
\left(  \bar{e}_L \gamma_\mu  C_A^+\nu 
-  \bar{e}_R  \gamma_\mu   C_A^-\nu  \right)  
\nonumber\\
&-&  
\bar{p}n \left( 
 \bar{e}_R C_S^+ \nu +  \bar{e}_L C_S^-\nu  \right)
- \frac{1}{2}\bar{p}\sigma^{\mu\nu} n 
\left(  \bar{e}_R \sigma_{\mu\nu}C_T^+ \nu
+  \bar{e}_L \sigma_{\mu\nu} C_T^-\nu  \right) 
\nonumber\\
& + &  \bar{p} \gamma_5 n  \left(  \bar{e}_RC_P^+\nu 
-   \bar{e}_LC_P^-\nu  \right)
+ \hc   
\eea
In our analysis, we will focus on the couplings to neutrinos with right-handed chirality, corresponding to the couplings, $C_{V,A,S,P,T}^-$ in the above Lagrangian, or $C_{\rm \Gamma A R}^{(6)}$ of Eq.\ \eqref{6final}. Given the matching discussed in App.\ \ref{app:match}, the relevant couplings are all induced by interactions that involve sterile neutrinos~\footnote{Note that these interactions can also be induced by operators involving left-handed neutrinos once one considers operators of dimension seven, since $\nu_L^c$ is right handed. However, such operators violate lepton number by two units and are already stringently constrained by $0\nu\bt\bt$, leading to $\Lambda>10$ TeV \cite{Cirigliano:2017djv}. This implies their effects are unlikely to be measured in  $\bt$ experiments, given the sensitivity to the $C_i^-$ couplings.}.

From the above Lagrangian one can already see the form of the contributions to single-$\bt$ decay and $0\nu\bt\bt$. The contributions to $\bt$ decay take the form $\Gamma_\bt\sim\sum_{\al,\bt}\sum_{i=1}^N\left[C_\al\right]_{ei} \left[C_\bt\right]^*_{ei}\Gamma_{\al\bt}(m_i)$, where $\Gamma_{\al\bt}(m_i)$ depends on the Wilson coefficients, $C_{\al,\bt}$, under consideration and is nearly independent of $m_i$ for small neutrino masses, $m_i\ll Q$. In this case, the contributions through the SM charged current reduce to $\Gamma_\bt^{\rm SM}\sim\sum_{i=1}^N\big|U_{ei} \big|^2\Gamma_{}(m_i)\simeq \Gamma_{}(0)$ and are independent of neutrinos masses and mixings. For large $m_i$ the partial rates $\Gamma_{\al\bt}(m_i)$ depend more sensitively on the neutrino masses until they vanish for $m_i>Q$, as the decay to $\nu_i$ becomes energetically disallowed. This effect can be searched for experimentally by looking for kinks in the electron spectrum \cite{Shrock:1980vy,Bryman:2019bjg}.

Instead, the $0\nu\bt\bt$ decay rate is an LNV observable and therefore requires an explicit insertion of the neutrino mass leading to, $\Gamma_{0\nu\bt\bt}\sim\big|\sum_{\al,\bt}\sum_{i=1}^N \left[C_\al\right]_{ei} m_i \left[C_\bt\right]_{ei}\mathcal A_{\al\bt}(m_i)\big|^2$. Similar to $\bt$ decay, the amplitudes $\mathcal A_{\al\bt}(m_i)$ depend on the Wilson coefficients under consideration and are nearly independent of $m_i$ for small neutrino masses, $m_i\ll m_\pi$. For large neutrino masses, the amplitudes scale as $\mathcal A_{\al\bt}(m_i)\sim 1/m_i^2$, with a more complicated dependence in between, see Sect.\ \ref{0vbb} and Ref.\ \cite{Dekens:2020ttz} for more details.

The relation between the parameters in Eq.\ \eqref{eq:TH_Lleeyang} and those in Table \ref{tab:translation} is given by~\cite{Gonzalez-Alonso:2018omy}
\bea
\label{eq:TH_LYtoRWEFT}
C_V^+ & = & {V_{ud} \over  v^2} g_V \sqrt{1 + \Delta_R^V} \big ( 1+ \epsilon_L + \epsilon_R \big ) , 
\qquad 
C_V^-  =  {V_{ud} \over v^2} g_V  \sqrt{1 + \Delta_R^V}  \big ( \tilde \epsilon_L + \tilde \epsilon_R \big ) , 
\nnl 
C_A^+ & = & - {V_{ud} \over v^2} g_A  \sqrt{1 + \Delta_R^A}  \big ( 1+ \epsilon_L - \epsilon_R \big ) , 
\qquad 
C_A^-  =  {V_{ud} \over v^2} g_A  \sqrt{1 + \Delta_R^A}  \big ( \tilde \epsilon_L - \tilde \epsilon_R \big ) , 
\nnl
C_T^+ & = & {V_{ud} \over v^2} g_T \epsilon_T  , 
\qquad 
C_T^-  =  {V_{ud} \over v^2} g_T \tilde \epsilon_T, 
\nnl
C_S^+ & = & {V_{ud} \over v^2} g_S \epsilon_S  , 
\qquad 
C_S^-  =  {V_{ud} \over  v^2} g_S \tilde \epsilon_S,  
\nnl
C_P^+ & = & {V_{ud} \over v^2} g_P \epsilon_P  , 
\qquad 
C_P^-  =  - {V_{ud} \over  v^2} g_P \tilde \epsilon_P,  
\eea 
where $g_{V,A,S,P,T}$ are the vector, axial, scalar, pseudoscalar, and tensor charges of the nucleon~\cite{Gonzalez-Alonso:2018omy}, which can be determined using lattice calculations. 
Here,  $g_V = 1$ up to (negligible) quadratic corrections in isospin-symmetry breaking~\cite{Ademollo:1964sr} and we use the  FLAG'19 averages~\cite{Aoki:2019cca} for the axial, scalar and tensor charges: $g_A = 1.251(33)$, $g_S = 1.022(100)$ and $g_T = 0.989(33)$~\cite{Chang:2018uxx,Gupta:2018qil} (see also Ref.~\cite{Gonzalez-Alonso:2013ura}). 
Although the pseudoscalar charge is enhanced by the pion pole, namely $g_P=349(9)$~\cite{Gonzalez-Alonso:2013ura}, the suppression of the pseudoscalar contributions to single $\beta$-decay is larger such that $\beta$-decay experiments do not significantly constrain $C_P^-$. 

The matching in Eq.\ \ref{eq:TH_LYtoRWEFT} includes the short-distance (inner) radiative corrections $\Delta_R^V$ and $\Delta_R^A$. 
Especially the former is important, because it is necessary to extract $V_{ud}$ from nuclear data. 
Four recent calculations of this quantity are available~\cite{Seng:2018yzq,Czarnecki:2019mwq,Seng:2020wjq,Hayen:2020cxh}, which all agree within $1\sigma$. In  this analysis we use the Seng et al.\ evaluation, $\Delta_R^V = 0.02467(22)$~\cite{Seng:2018yzq}, which has the smallest uncertainty. 
Given  the assumption about the reality of $\epsilon_X$ and $\tilde \epsilon_X$, all $C_X^\pm$ are then also real.

\begin{table}[t!]
\begin{center}
\begin{tabular}{|c|c|}
	\hline
	$C^{(6)}_{\textrm{VLL}} = -2 V_{ud} (1 + \eL)$ & $ C^{(6)}_{\textrm{VLR}} = -2 V_{ud} \, \teL$ \\ [2ex]
	\hline
	$C^{(6)}_{\textrm{VRL}} = -2 V_{ud} \, \eR$ & $ C^{(6)}_{\textrm{VRR}} = -2 V_{ud} \, \teR$ \\ [2ex]
	\hline
	$C^{(6)}_{\textrm{SLL}} = - V_{ud} (\eS + \eP)$ & $C^{(6)}_{\textrm{SLR}} = - V_{ud} (\teS + \teP)$ \\ [2ex]
	\hline
	$C^{(6)}_{\textrm{SRL}} = - V_{ud} (\eS - \eP)$ & $C^{(6)}_{\textrm{SRR}} = - V_{ud} (\teS - \teP)$ \\ [2ex]
	\hline
	$C^{(6)}_{\textrm{TLL}} = -\12 V_{ud} \, \eT$ & $C^{(6)}_{\textrm{TRR}} = -\12 V_{ud} \, \teT$ \\ [2ex]
	\hline
\end{tabular}
\end{center}\label{tab:translation}
\caption{Translation to the notation of \cite{Falkowski:2020pma}.}
\end{table}

\section{Single- and neutrinoless double-$\bt$ decay observables}

\subsection{Single $\bt$-decay}\label{FGN}
As mentioned the interactions in Eq.\ \eqref{eq:TH_Lleeyang} give tree-level corrections to the SM charged current which can be probed in $\bt$ decays.
We will follow the analysis of Ref.\ \cite{Falkowski:2020pma} to constrain the couplings in  Eq.\ \eqref{eq:TH_Lleeyang}~\footnote{We thank Adam Falkowski and Mart\'in Gonz\'alez-Alonso for providing us with the $\chi^2$ function used in the analysis of Ref.\  \cite{Falkowski:2020pma}}. Here we briefly summarize this analysis by FGN and discuss some of the details, related to theory uncertainties, of the fit.

Since the observables considered by FGN include the $\bt$ decays of the neutron as well as nuclei, the theoretical expressions depend of the NMEs in addition to the hadronic matrix elements appearing in Eq.\ \eqref{eq:TH_LYtoRWEFT}.
 At leading order, nuclear effects are encapsulated in the so-called Fermi (F) and Gamow-Teller (GT) matrix elements, $M_{F,GT}$. Leading-order expressions for beta decay observables in terms of the Lee-Yang Wilson coefficients $C_X^\pm$ can be found in Refs.~\cite{Jackson:1957zz,EBEL1957213}. 
However, given the experimental precision, subleading effects such as weak-magnetism and long-distance electromagnetic corrections have to be included in the SM terms. These small contributions can be calculated with high accuracy for the transitions that are included in this work~\cite{Holstein:1974zf,Cirigliano:2013xha,Hayen:2017pwg,Hayen:2020nej}.

Apart from the corrections that can be accurately determined, there are several places where theoretical uncertainties are sizable. In particular, the ``polluted'' mixing parameters, $\tilde \rho_i$, of \cite{Falkowski:2020pma} involve ratios of GT to F matrix elements which have not been calculated to sufficient precision. They are therefore treated as nuisance parameters and are marginalized over in the fit. Furthermore, theoretical uncertainties related to the nuclei-dependent parts of the radiative corrections is taken into account by including two additional nuisance parameters, $\eta_{2,3}$   \cite{Falkowski:2020pma}.

In Ref.~\cite{Falkowski:2020pma}, the best-fit results are obtained by marginalizing over nuisance parameters and other Wilson coefficients.
In this study, depending on the scenario under consideration, we do not always marginalize over all the Wilson coefficients.
In some cases we keep certain Wilson coefficients fixed, in order to be able to compare directly with the $0\nu\bt\bt$ limits. As a result, the $\bt$ limits we present are not always identical to those presented in Ref.\ \cite{Falkowski:2020pma}. We do always marginalize over all the nuisance parameters, even when only turning on one or two Wilson coefficients at a time.

Finally, although FGN neglected all neutrino masses, a complete analysis would in principle include the effects that the neutrino masses, $m_i$, have on single-$\bt$ observables. Instead of taking into account the full $m_i$ dependence of all $\bt$-decay observables in the fit, we will restrict ourselves to a range of $m_i$ for which the FGN analysis is still applicable. 
We estimate an upper bound on this range by considering the ways in which $\bt$ observables will receive corrections due to nonzero $m_i$. First, the contributions induced by the SM charged current are proportional to $\sum_{i=1}^N |U_{ei}|^2 = 1$ for negligible neutrino masses, with corrections of the form~\footnote{An explicit calculation of the $\bt$ decay rate confirms this scaling.} $\sum_{i=1}^N |U_{ei}|^2 \frac{m_i^2}{Q^2}$. Here, the $Q$ values of the isotopes that appear in the fit are in the MeV range, $Q\gtrsim 0.7$ MeV. 
Combining this with the upper limit on the mixing parameter $U_{e4}$, roughly $|U_{e4}|^2 < 10^{-3}$ for $m_4 = 100$ keV from the $\beta$-decay spectrum of ${}^{35}$S \cite{Shrock:1980vy,Holzschuh:2000nj,Bolton:2019pcu}, the corrections to the $\bt$ decay rates due to a fourth neutrino are expected to be below the $\Or(10^{-4})$ level, for $m_4\lesssim 100$ keV. In addition, the contributions of dimension-six operators will receive corrections $\sim \frac{m_i^2}{Q^2} |v^2 C_\al^-|^2$, or, allowing for interference, $\sim  \frac{m_i}{Q} |U_{e4}\,v^2 C_\al^-|$. Since the higher-dimensional operators will be suppressed compared to the SM, $v^2 C_\al^-\lesssim 0.1$, we expect these corrections to contribute at most at the $\Or(10^{-4})$ level. The observables in the FGN analysis, such as phase-space corrected lifetimes ($\mathcal F$t values) and $\beta$-decay asymmetries, have $\mathcal O(10^{-3})$ uncertainties and are thus not sensitive to neutrino-mass corrections for  $m_4\lesssim100$ keV.

\subsection{Neutrinoless double beta decay }\label{0vbb}
We now turn to the calculation of $0\nu\beta\beta$ decay rates of various nuclear isotopes. We closely follow the framework of Ref.~\cite{Dekens:2020ttz}. Our starting point is the following effective Lagrangian of Eq.\ \eqref{6final}
\bea
\mathcal L^{(6)}& =& \frac{2 G_F}{\sqrt{2}} \Bigg\{ 
  \bar u_L \gamma^\mu d_L \left[   \bar e_{L}  \gamma_\mu  C^{(6)}_{\textrm{VLL}}  \,  \nu + \bar e_{R}  \gamma_\mu C^{(6)}_{\textrm{VLR}} \,  \nu \right]+
  \bar u_R \gamma^\mu d_R \left[\bar e_{R}\,  \gamma_\mu  C^{(6)}_{\textrm{VRR}} \,\nu \right]\nn\\
& & +
  \bar u_L  d_R \left[ \bar e_{L}\, C^{(6)}_{ \textrm{SRR}}  \nu  \right]+ 
  \bar u_R  d_L \left[ \bar e_{L} \, C^{(6)}_{ \textrm{SLR}}    \nu \right]+  \bar u_L \sigma^{\mu\nu} d_R\,  \bar e_{L}  \sigma_{\mu\nu} C^{(6)}_{ \textrm{TRR}} \, \nu \Bigg\}  +{\rm h.c.}\eea
  This Lagrangian is written in the neutrino mass basis where $\nu$ denotes a $3+n$ column vector of neutrino mass eigenstates. The Lagrangian is defined at a scale $\mu=2$ GeV and consists of operators that can be induced by operators of the dimension-six $\nu$SMEFT Lagrangian involving one $\nu_R$ field. The only exception is the $C^{(6)}_{\textrm{VLL}}$ term which also includes the SM charged weak interaction of the active neutrinos. 
  
 We consider a simplified scenario where $n=1$ implying a single sterile neutrino with arbitrary mass $m_4$. While a pure $3+1$ model (without dimension-five operators) would lead to two massless neutrinos, we assume some mass mechanism for the active neutrinos for instance through additional decoupled sterile neutrinos. As discussed above, we focus on $m_4$ masses well below the MeV scale when comparing to single-$\beta$ experiments, where (in most cases) such neutrinos can be treated as massless~\footnote{Exceptions are experiments, such as KATRIN, aiming to directly measure the neutrino mass \cite{Riis:2010zm,Giunti:2019fcj,Aker:2020vrf}.}, while we consider larger masses when investigating the reach of $0\nu\beta\beta$ experiments.  
 
 Under these assumptions it is possible to write down a relatively simple expression for the $0\nu\beta\beta$ decay rate
 \bea\label{eq:T1/2}
\left(T^{0\nu}_{1/2}\right)^{-1} &=& g_A^4 \bigg\{ G_{01} \, \left( |\mathcal A_{L}|\sq + |\mathcal A_{R}|\sq \right)
- 2 (G_{01} - G_{04}) \textrm{Re} \mathcal A_{L}^* \mathcal A_{R} 
\nn\\
&&+ G_{09}\, |\mathcal A_{M}|\sq + G_{06}\, {\rm Re}\left[ (\mathcal A_{L} - \mathcal A_{R} )\mathcal A_{M}^*\right] \bigg\}\,,
\eea
in terms of electron phase-space factors, $G_{0i}$, and three so-called subamplitudes $\mathcal A_{L,R,M}$. The $G_{0i}$ phase space factors are defined in Ref.~\cite{Cirigliano:2017djv} and have been calculated in the literature \cite{Kotila:2012zza,Stefanik:2015twa,Horoi:2017gmj}. They include Coulomb corrections between the outgoing electrons and the nucleus and the (partial) screening effect of the nuclear charge by the surrounding electron cloud. Numerical values are given in Table~\ref{Tab:phasespace}.

\begin{table}
\center
\begin{tabular}{|c|cccc|}
\hline
\hline
\cite{Horoi:2017gmj}	    & $^{76}$Ge & $^{82}$Se & $^{130}$Te & $^{136}$Xe \\ 

\hline
$G_{01}$    & 0.22 & 1. & 1.4 & 1.5 \\
$G_{04}$    & 0.19 & 0.86 & 1.1 & 1.2 \\
$G_{06}$    & 0.33 & 1.1 & 1.7 & 1.8 \\
$G_{09}$    & 0.48 & 2. & 2.8 & 2.8 \\\hline
\hline
$Q/{\rm MeV} $ \cite{Stoica:2013lka} & 2.04& 3.0&2.5 & 2.5 \\
\hline\hline
\end{tabular}
\caption{Phase space factors in units of $10^{-14}$ yr$^{-1}$ obtained in Ref.~\cite{Horoi:2017gmj}. The last row shows the $Q$ value of \NLDBD\ for various isotopes, where $Q = M_i - M_f -2m_e$.}
\label{Tab:phasespace}
\end{table}

The subamplitudes depend on the underlying LNV mechanism (in our case, the Majorana mass terms and the non-standard neutrino couplings) as well as hadronic and nuclear matrix elements. Each subamplitude is written as a sum over neutrino mass eigenstates 
\begin{equation}\label{eq:subamps}
 \mathcal A_{L,R,M} = \sum_{i=1}^{3+n} \mathcal A_{L,R,M}(m_i)\,.
\end{equation}
Strictly speaking this description is only valid if $m_i < 1$\,GeV or so. Heavier neutrinos must be integrated out leading to effective dimension-nine interactions which give additional contributions in Eq.\ \eqref{eq:subamps}. In what follows we capture these effects by demanding that the $m_i$-dependent nuclear and hadronic matrix elements have the correct $m_i$ behavior to reproduce the amplitudes that would result from integrating out the neutrinos at the quark level, in the $m_i\gg 1$ GeV region. As we focus on neutrino masses smaller than a GeV or so, we relegate a discussion of the procedure to App.\ \ref{app:double} and refer to Ref.\ \cite{Dekens:2020ttz} for more details.

 The first subamplitude is given by 
\begin{eqnarray}\label{Anu}
\mathcal A_L(m_i) &=&  -\frac{m_i}{4 m_e} \Bigg\{ \left[\mathcal M_V(m_i) + \mathcal M_A(m_i)\right] \left(C^{(6)}_{\rm VLL} \right)^2_{ei}  \nn \\
&& + \mathcal M_{PS}(m_i) \bigg[  \frac{B^2}{m_\pi^2}  \left( C^{(6)}_{\rm SRR} -C^{(6)}_{\rm SLR} \right)_{ei}
  - 2  \frac{B}{m_i} \left(C^{(6)}_{\rm VLL}  \right)_{ei}   
\bigg] \left(C^{(6)}_{\rm SRR} - C^{(6)}_{\rm SLR}\right)_{ei} \nn\\
 & & +  \mathcal M_S(m_i) \left( C^{(6)}_{\rm SRR} + C^{(6)}_{\rm SLR} \right)^2_{ei} - \mathcal M_T(m_i) \left( C^{(6)}_{\rm TRR} \right)^2_{ei} 
 \nn \\
& & +  \frac{m_N}{m_i} \mathcal M_{TV}(m_i) \left(C^{(6)}_{\rm VLL} \right)_{ei} \left(C^{(6)}_{\rm TRR}\right)_{ei} 
  \Bigg\}+\mathcal A_L^{(\nu)}(m_i)\,,    
 \end{eqnarray}
 in terms of the low-energy constant (LEC) $B \simeq 2.7$ GeV, $m_N$ and $m_\pi$ the nucleon and pion mass, respectively, and various combinations of neutrino-mass-dependent NMEs $\mathcal M_{I}(m_i)$ which are discussed below. Most terms in the expression for $\mathcal A_L(m_i)$ have an explicit $m_i$ dependence accounting for the required lepton number violation. Two interference terms, linear in $\left(C^{(6)}_{\rm VLL}  \right)_{ei}$, seem to violate this scaling. However,  in order for the product with $\left(C^{(6)}_{\rm SRR} - C^{(6)}_{\rm SLR}\right)_{ei}$ or $\left(C^{(6)}_{\rm TRR}\right)_{ei}$ to be non-vanishing, $C^{(6)}_{\rm VLL} $ requires an insertion of $M_R$, see the last term in Eq.\ \eqref{match6LNC}, so that the relevant terms $\left(C^{(6)}_{\rm VLL}  \right)_{ei} \sim m_i v /\Lambda^2$ scale linearly in the Majorana mass. These interference terms are then effectively suppressed by $B/v \sim m_N/v= \mathcal O(10^{-2})$ and can be neglected.
 
 The term $\mathcal A_L^{(\nu)}(m_i)$ describes contributions from the exchange of hard neutrinos, virtual neutrinos with momenta $\gtrsim$ GeV. Such contributions lead to $\mathcal O(1)$ corrections to the amplitudes, but in most cases depend on QCD matrix elements that are not well known. For our numerical analysis we have used lattice QCD determinations \cite{Nicholson:2018mwc} when possible, while the remaining matrix elements are estimated using naive dimensional analysis (NDA), see App.\ \ref{app:double} and Ref.~\cite{Dekens:2020ttz} for more details.

Similarly
\begin{eqnarray}\label{AR}
\mathcal A_R(m_i) &=&  -\frac{m_i}{4 m_e} \Bigg\{ \mathcal M_V(m_i) \left( C^{(6)}_{\rm VRR} + C^{(6)}_{\rm VLR} \right)^2_{ei} 
					+ \mathcal M_A(m_i) \left( C^{(6)}_{\rm VRR} - C^{(6)}_{\rm VLR} \right)^2_{ei}  \Bigg\}\nn\\
					&&+\mathcal A_R^{(\nu)}(m_i)\,,
\end{eqnarray}
while the final subamplitude contains only interference terms
\begin{eqnarray}\label{AM}
\mathcal A_M(m_i) &=& 
-\frac{m_i}{2 m_e} \Bigg\{  \frac{m_N}{m_i} \mathcal M_{VA}(m_i) \left( C^{(6)}_{\rm VLL}\right)_{ei}\left( C^{(6)}_{\rm VLR}\right)_{ei} \nn \\
& &  +\frac{1}{2}  \mathcal M_S(m_i) \frac{g_V}{g_S} \left( C^{(6)}_{\rm VRR} + C^{(6)}_{\rm VLR} \right)_{ei} \left( C^{(6)}_{\rm SRR} + C^{(6)}_{\rm SLR} \right)_{ei}  \nn \\
& & + \mathcal M_{T A}(m_i) 
 \left( C^{(6)}_{\rm VRR} - C^{(6)}_{\rm VLR} \right)_{ei} \left( C^{(6)}_{\rm TRR} \right)_{ei}  \bigg\} +\mathcal A_M^{(\nu)}(m_i)  \,.
\end{eqnarray}
Here the $\mathcal A_{R,M}^{(\nu)}(m_i)$ subamplitudes are again due to the exchange of hard neutrinos, which we discuss in App.\ \ref{app:double}.

As mentioned, the $\mathcal M_{I}(m_i)$ are linear combinations of different NMEs which are discussed in App.\ \ref{app:double}. These elements in principle depend on the mass of the neutrino that is being exchanged. However, for neutrino masses below a few MeV, this mass dependence can be safely neglected. The mass-independent NMEs, $\mathcal M_{I}(0)$, have been calculated with various nuclear methods. A collection of results is given in App.\ \ref{app:double}, Table \ref{tab:comparison} for several isotopes. The differences between nuclear calculations lead to an additional $\mathcal O(1)$ uncertainty on the $0\nu\beta\beta$ decay results \cite{Engel:2016xgb}. 

For heavier neutrino masses, the mass dependence of the NMEs and hadronic low-energy constants cannot be neglected. We follow the approach of Ref.~\cite{Dekens:2020ttz} which interpolates between regions where $m_i \ll k_F \sim m_\pi$ where $k_F$ is the typical Fermi momentum in nuclei and $m_i > \Lambda_\chi \sim1$ GeV where neutrinos can be integrated out at the quark level leading to local LNV operators. In the small-mass regime, NMEs and LECs are essentially mass independent and the amplitude scales linearly with $m_i$. In the large-mass regime, the neutrino mass-dependence is captured by the Wilson coefficients of dimension-nine operators, $\sim (\bar ud)^2 \bar ee^c$, that are generated after integrating out the neutrinos. The Wilson coefficients then have the form $C^{(6)}_\al C^{(6)}_\bt/m_i$ so that the amplitude scales as $1/m_i$, while the NMEs and LECs are again mass independent. 
The mass dependence in the intermediate regime arises from two sources: 1) the neutrino propagator scales as $m_i / (\vec q^2 +m_i^2)$ instead of the Coulomb-like $m_i/|\vec q|$, 2) LECs associated to the exchange of hard virtual neutrinos pick up an intrinsic $m_i$ dependence.  Here we will apply the formulae that were constructed in Ref.~\cite{Dekens:2020ttz} by matching to results in the low- and high-mass regimes and  interpolating in between. These formulae thus have the largest uncertainty in the $m_i \sim \Lambda_\chi$ regime where the chiral and perturbative-QCD descriptions cannot be trusted, see App.\ \ref{app:double} for explicit expressions.

\section{Results}\label{results}

We now turn to the calculation of $0\nu\beta\beta$ constraints on the various effective interactions. We begin our analysis by considering a single $C^-_{X}$ coupling, while setting the remaining Wilson coefficients to their SM values. We compare the constraints from $0\nu\beta\beta$ decay against single-$\beta$ limits~\cite{Falkowski:2020pma}. The former are only valid for Majorana neutrinos. The latter are strictly speaking only valid in the limit of massless neutrinos but we will consider neutrino masses $m_{\nu_R}<100\,{\rm keV}$, where this is a reasonable approximation, see the discussion in Sect.\ \ref{FGN}. 
 \begin{figure}[t!]\begin{center}
\includegraphics[scale=0.55]{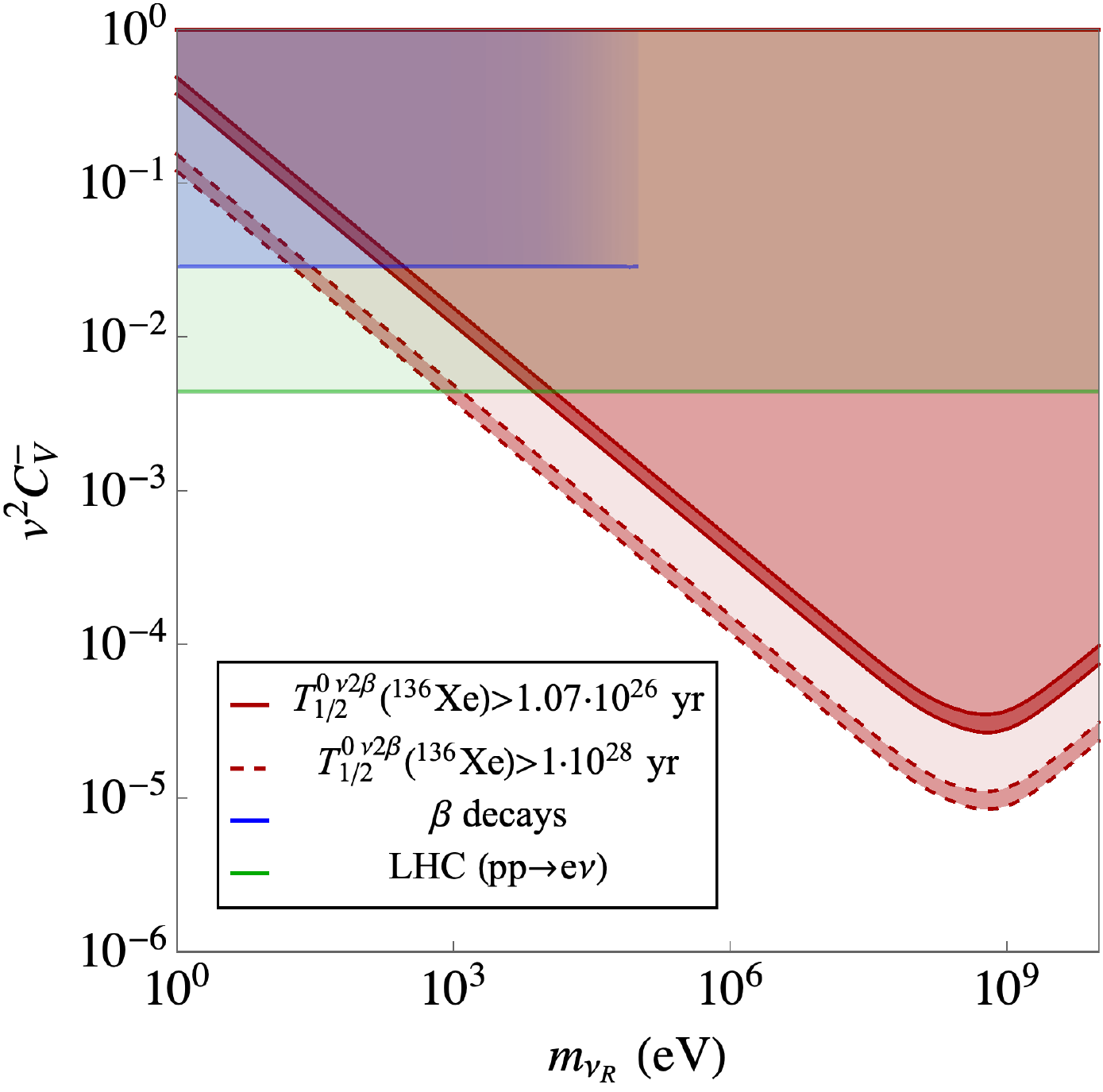}\hfill
\includegraphics[scale=0.55]{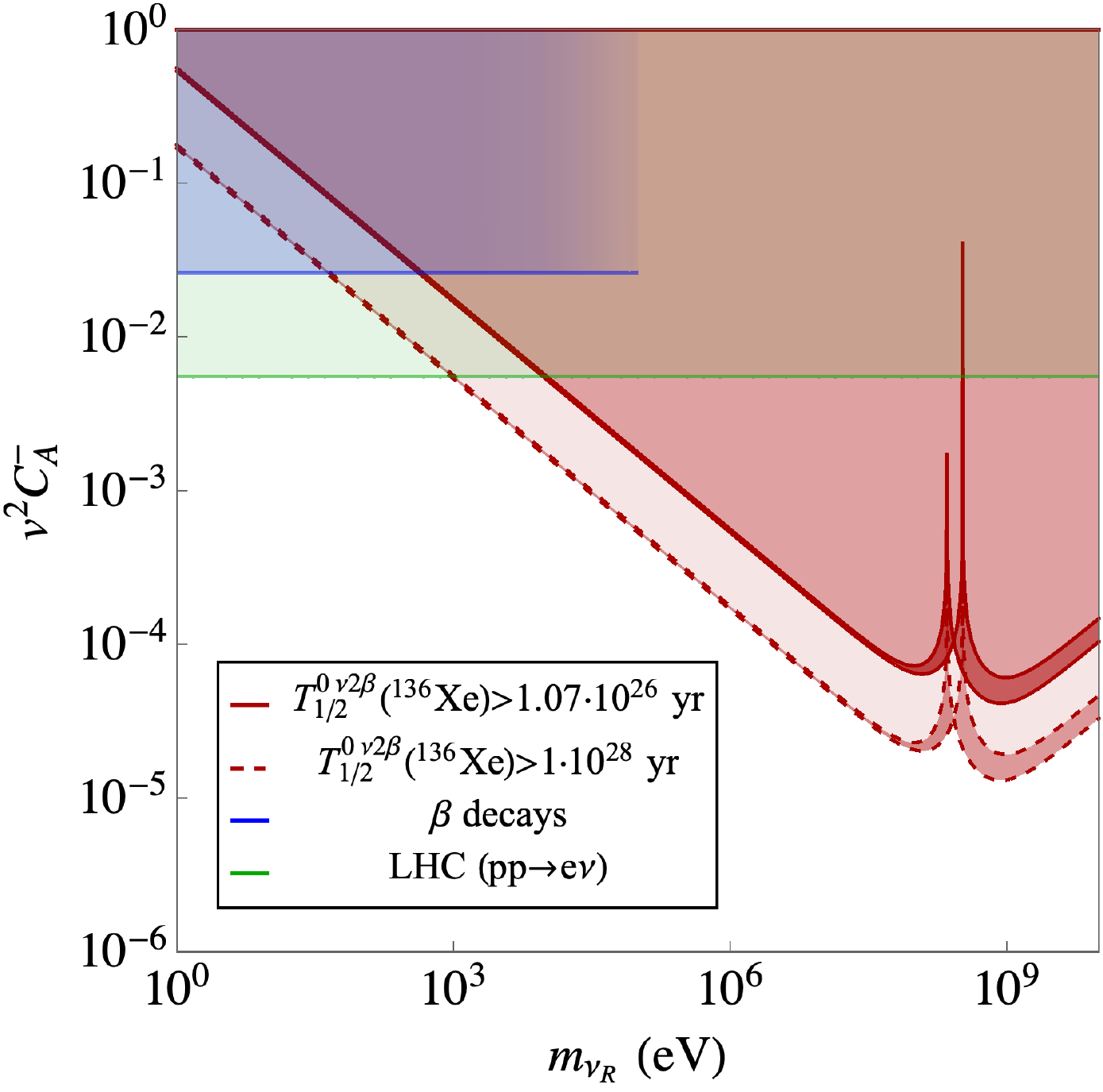}\\
\includegraphics[scale=.55]{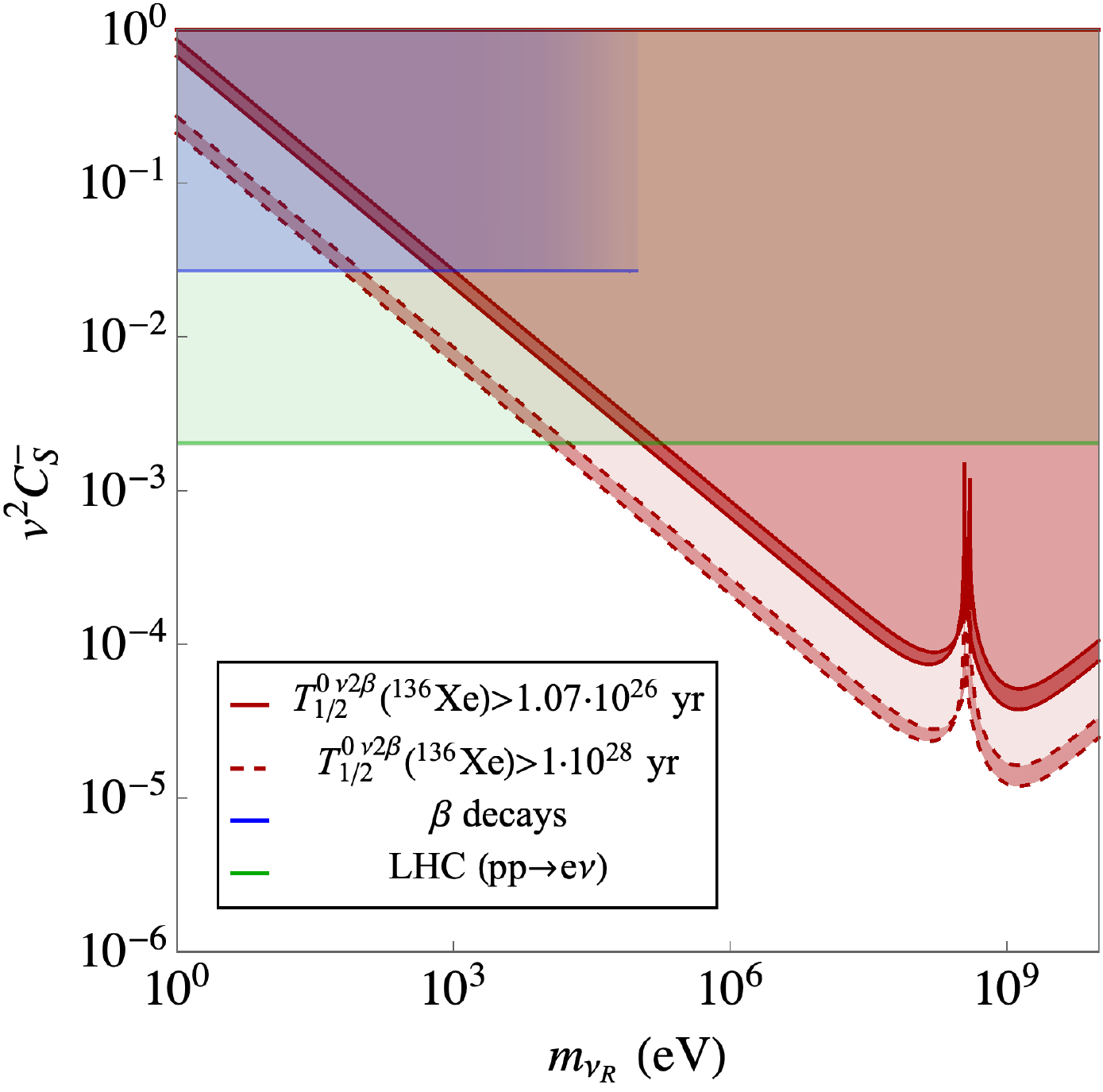}\hfill
\includegraphics[scale=0.55]{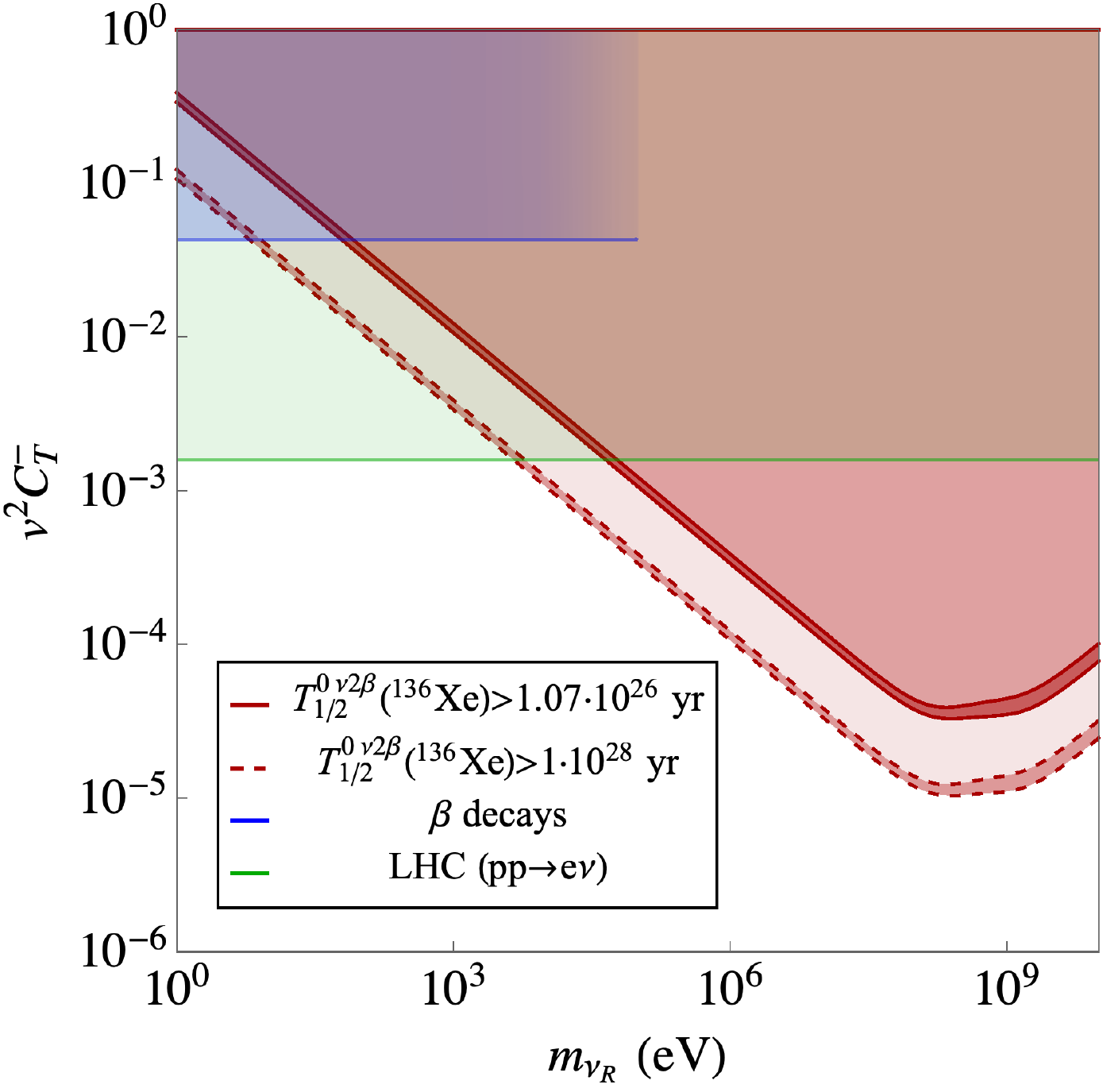}
\end{center}
\caption{
90\% C.L.\ limits from the $0\nu\bt\bt$ decay of $^{136}$Xe (red) compared to a fit of $\bt$ decays (blue) in the scenario where only one of the  $v^2 C^{-}_{i}$ has been turned on. The width of the $0\nu\bt\bt$ bands indicate the variation induced by using shell-model \cite{Menendez:2017fdf} or QRPA \cite{Hyvarinen:2015bda} NMEs.}
\label{fig:plots}
\end{figure}

To derive the contributions of the scenarios considered below to $0\nu\bt\bt$ we make several assumptions. We will consider a case with one additional neutrino, $n=1$, and assume that the non-standard interactions only couple to the sterile state, $i=4$ (this is equivalent to assuming there is no mixing between sterile and active neutrinos, $U_{4i}\propto \dt_{4i}$). In addition, we will not consider the `usual' mechanism for inducing $0\nu\bt\bt$ proportional to $m_{\bt\bt}= m_i U_{ei}^2$. Both simplifications neglect additional contributions to $0\nu\bt\bt$; while the neglected mixing effects can only contribute through interference terms and will tend to be suppressed (see Sect.\ \ref{0vbb}), the $\sim m_{\bt\bt}$ terms can in principle be sizable. The reason we nevertheless make these simplifications is that these effects depend on the details of the neutrino masses and mixings, determined by $M_{L,D,R}$ in Eq.\ \eqref{eq:numasses}, which can only be assessed in a full model of neutrino masses. Barring cancelations, these additional contributions will tend to increase the predicted $0\nu\bt\bt$ rate, thereby strengthening the limits presented here. Our approach is thus conservative.

\subsection{Single coupling analysis}
\begin{figure}[t!]\begin{center}
\includegraphics[scale=0.55]{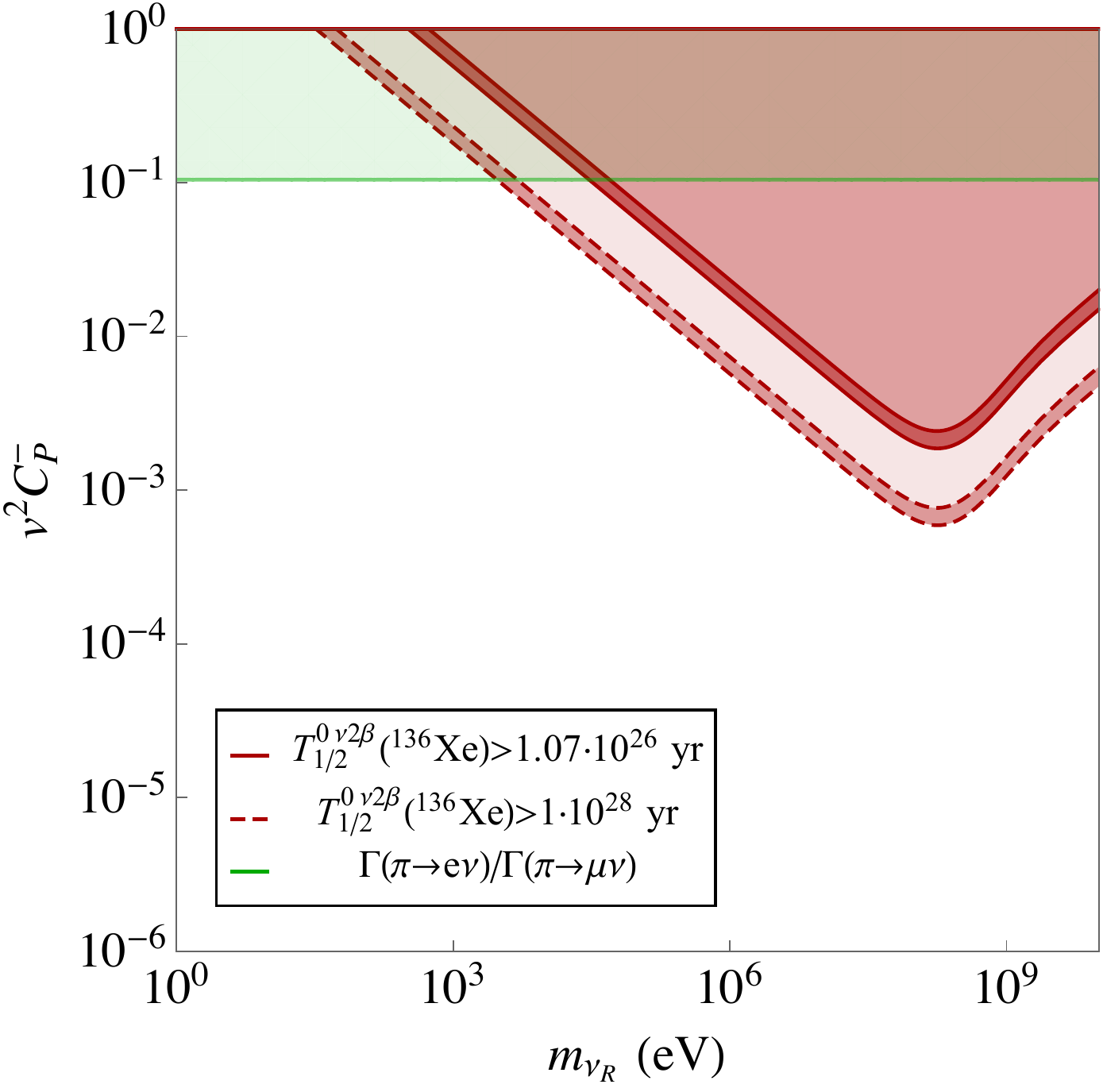}
\end{center}
     \caption{
Same as Fig.\ \ref{fig:plots} but for $C^-_P$.
}
       \label{fig:plots2}
\end{figure}

Our results are shown in Figs.~\ref{fig:plots} and \ref{fig:plots2} for the 5 different couplings. The plots depict the $m_4-v^2C^-_i$ plane, where the blue (red) shaded areas show the parameter space allowed by the single-$\bt$ (double-$\bt$) constraints.
To obtain the latter we assume there are no significant cancelations between the $\nu_R$ contributions and those proportional to $m_{\bt\bt}$. 
The width of the $0\nu\bt\bt$ bands show the variation induced by using shell-model \cite{Menendez:2017fdf} or quasi-particle random phase approximation (QRPA) \cite{Hyvarinen:2015bda} NMEs, illustrating the impact of the nuclear theory uncertainties. 
The panels do not show the effect of changes in the LECs discussed in App.\ \ref{app:hardnu}, significant uncertainties of which would lead to additional errors at the  $\Or(1)$ level.
For reference we also show the LHC  limits and, for $C_P^-$, the constraint from $\Gamma(\pi\to e\nu)/\Gamma(\pi\to \mu\nu)$   in green \cite{Gonzalez-Alonso:2013uqa,Gupta:2018qil}. The collider limits were derived from $pp\to e +{\rm MET}+X$ based on $20(36)\, {\rm fm}^{-1}$ data recorded at $\sqrt{s} = 8(13)$ TeV for the $C^-_{V,A}$ ($C_{S,T}^-$) couplings.
In all cases the present KAMLAND-Zen limit are sensitive to couplings $v^2C^{-}_{i} <1$ for sterile masses larger than about $1$ eV. The $0\nu\beta\beta$ constraints become competitive with those from single-$\beta$ decay for masses larger than $\{180, 460, 610, 60\}$ eV in the case of $\{C_V^-,\,C_A^-,\,C_S^-,\,C_T^-\}$, respectively. 
Thus, if a nonzero value of $v^2C_T^-$ at the $\mathcal O(0.01)$ level would be confirmed in $\beta$-decay experiments~\footnote{The one-coupling fit roughly prefers nonzero values $|v^2C_T^-|\in [0.01,\,0.03]$ at $1\sigma$. The significance grows when allowing all Wilson coefficients to be nonzero, see Fig.\ \ref{fig:2Dmarg}.}, the present $0\nu\beta\beta$ limits imply that neutrinos are either Dirac particles or that $m_4 < 60$ eV. 
%The $0\nu\bt\bt$ limits are subject to $\Or(1)$ uncertainties due to theory errors related to the nuclear and hadronic matrix elements. Here we have used shell model NMEs \cite{Menendez:2017fdf} for concreteness, but the difference with other NMEs are hard to see on the applied logarithmic scale. 

The $0\nu\beta\beta$ decay bounds follow a slope $(C^{-}_{i})^2  \sim 1/ m_4$ for $m_4 < \mathcal O(100)$ MeV. This scaling changes to $(C^{-}_{i})^2  \sim m_4$ for $m_4 > \mathcal O(2)$ GeV, with a more complicated mass dependence in between. 
The exact location of the peak sensitivity depends on the coupling under consideration because of non-trivial mass dependence of nuclear and QCD matrix elements. 
To get an idea of the sensitivity to BSM scales that are probed we can assume a typical scaling $C^{-}_{i} = 1/\Lambda^2$. The constraints on $C^{-}_{V}$ then imply a BSM scale between $\Lambda >0.7 $ TeV for $m_4 = 10$ eV to a maximum sensitivity of $\Lambda >47 $ TeV for $m_4 = 0.5$ GeV. These sensitivities should be compared to $\Lambda>1.5$ TeV in the case of single-$\bt$ decay. 
The dips that appear around $m_4\sim$ GeV
for $C^-_{A}$ and $C^-_{S}$ arise from the fact that the $0\nu\bt\bt$ decay rate vanishes. In these cases, for our particular choice of hadronic and nuclear matrix elements, the long-distance contributions cancel the short-distance terms. At this point, higher-order corrections should be included and our calculations are not reliable. Considering that this happens only for a tiny range of neutrino masses we do not pursue this further. 
Finally, we denote the prospective limit, $T_{1/2}^{0\nu\bt\bt}>10^{28}$yr from EXO-200 \cite{Anton:2019wmi} by the dashed red lines. As this limit is a factor $\sim100$ stronger than the current constraint, it leads to an improvement of $\sim\sqrt{10}$ on the $C^-_i$ constraints.

The collider limits shown in Figs.~\ref{fig:plots} and \ref{fig:plots2} tend to be stronger than the single-$\bt$ constraints by a factor of a few to an order of magnitude. However, the LHC limits from $pp\to e\nu$  only apply as long as the BSM physics responsible for the dimension-six interactions can still be treated as a contact interaction at high energies, i.e.\ as long as, $\Lambda\gg\sqrt{s}\sim$ TeV. Instead, the limits from single- and double-$\bt$ decay are subject to a milder assumption, $\Lambda\gg$ GeV. In any case, if future LHC analyses find hints for interactions with light right-handed neutrinos, future $0\nu\beta\beta$ experiments should see a signal if the neutrinos are Majorana and have masses above roughly $1$ keV. 

Additional constraints  on $C_X^-$ arise from a careful analysis of the electron spectrum in neutrinoful double-$\beta$ decay \cite{Bolton:2020ncv,Agostini:2020cpz}. In particular, the angular distribution of outgoing electrons in ${}^{100}$Mo double-$\beta$ decay \cite{NEMO-3:2019gwo}, limits $\frac{v^2}{2}\left|\frac{C_{V}^-}{g_V}+\frac{C_{A}^-}{g_A}\right| \leq 0.03$ for $m_{\nu_R} < 0.1$ MeV comparable to the single-$\beta$ decay bounds. Future experiments are expected to improve this to the $0.01$ level \cite{Bolton:2020ncv}. We have not indicated these bounds in the plot. 

As seen from Fig.\ \ref{fig:plots2}, single-$\beta$ decay experiments are not sensitive to $C_P^-$ because of a $Q^2/m_\pi^2$ suppression of the amplitude where $Q$ denotes the $Q$-value of the experiment. However, the ratio $\Gamma(\pi\to e\nu)/\Gamma(\pi\to \mu\nu)$ is rather sensitive to this coupling, leading to a stringent constraint that is overtaken by the $0\nu\bt\bt$ limit for $m_4\gtrsim 30$ keV.

\subsection{Multi-coupling analysis}
\begin{figure}[t!]\begin{center}
\includegraphics[scale=0.55]{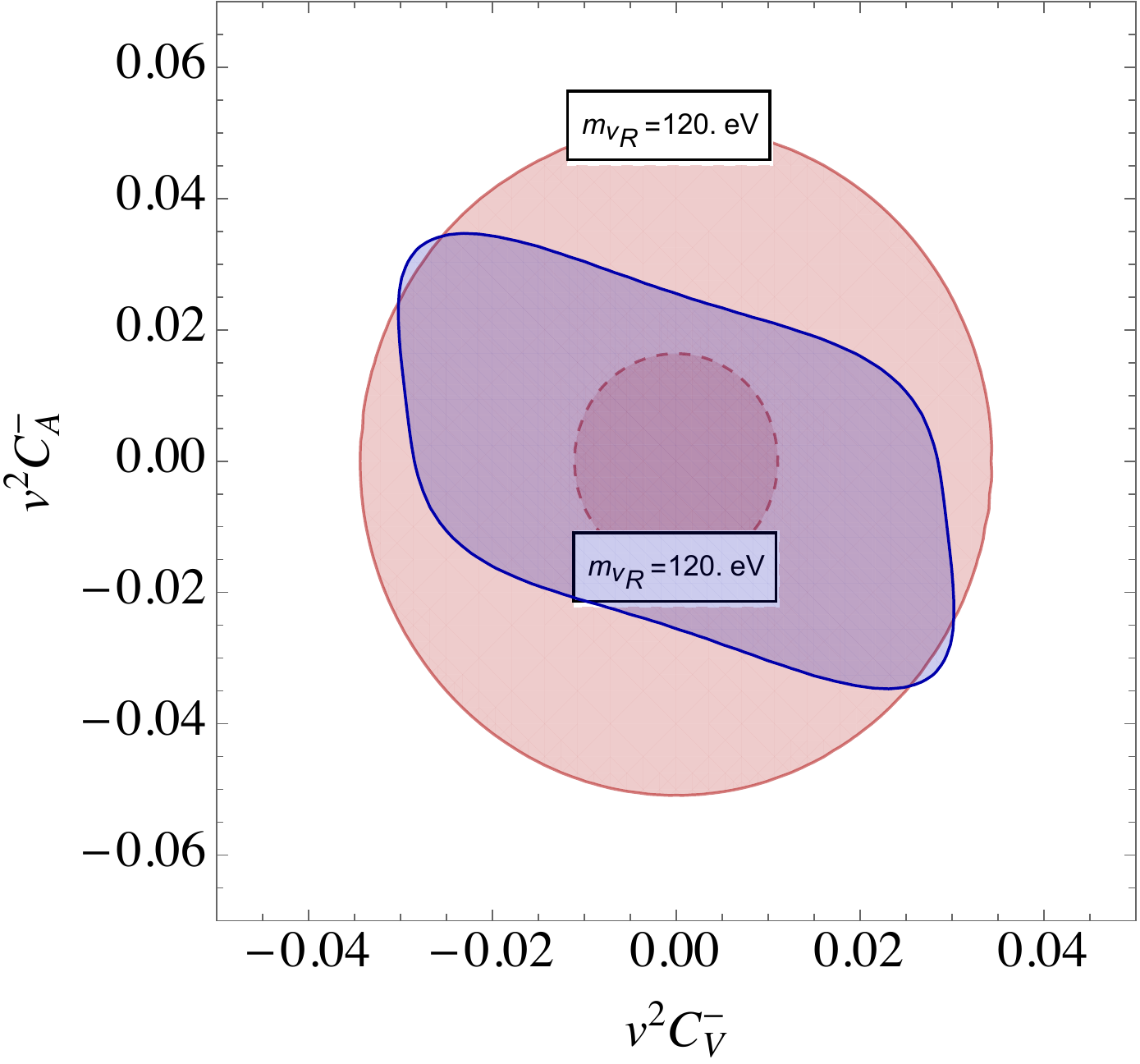}\hfill
\includegraphics[scale=0.55]{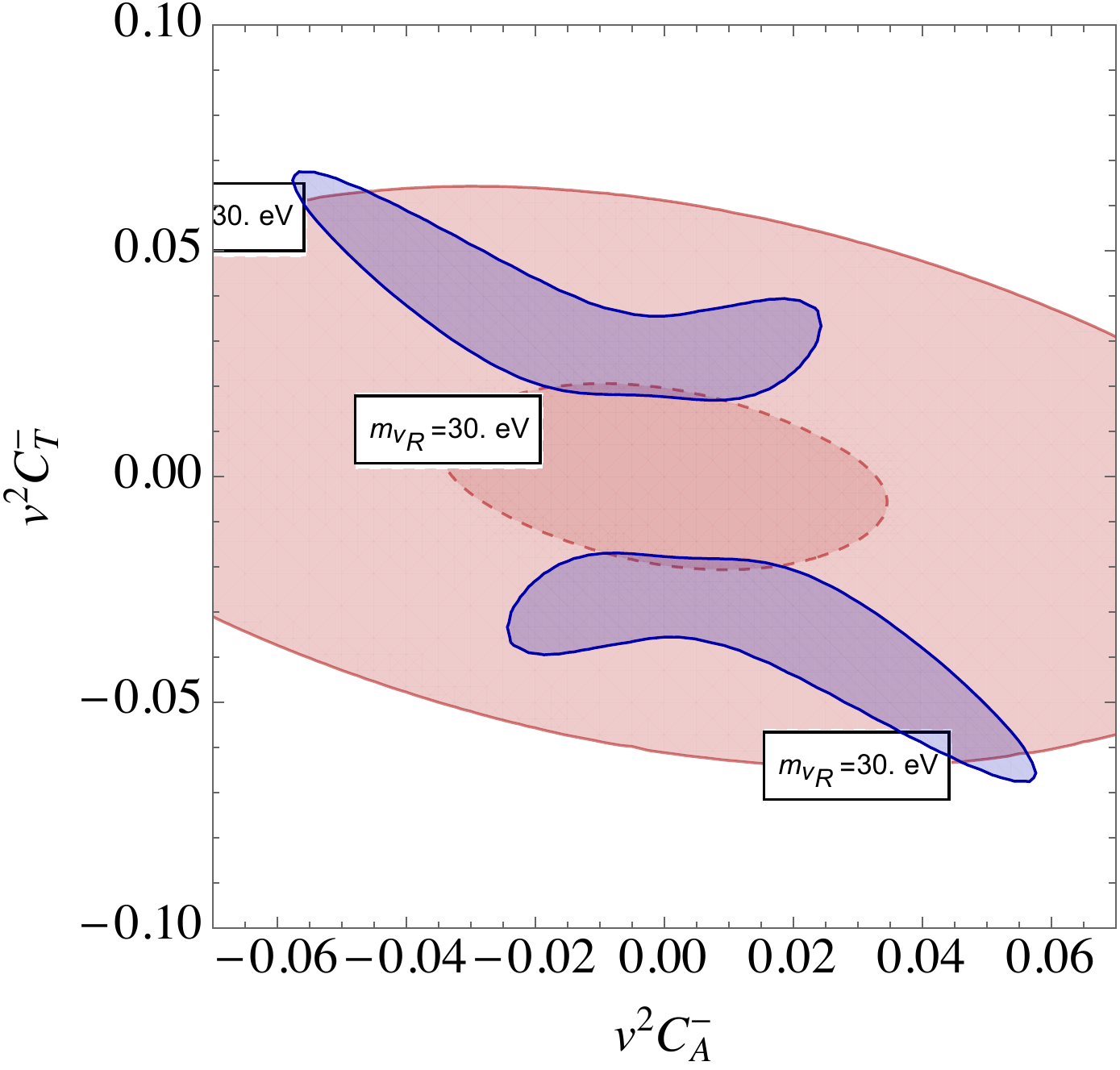}
\end{center}
     \caption{
Exclusion plots showing $\Delta \chi^2 = 2.706$ contours in the $v^2C_V^-$-$v^2C_A^-$ and $v^2C_A^-$-$v^2C_T^-$ planes in the left and right panels, respectively. The blue (red) regions are those allowed by single- (double-)$\bt$ decay constraints. In both cases we turned on only two Wilson coefficients at a time, setting the rest to their SM values.
}
       \label{fig:2Dplots}
\end{figure}

We now turn to a scenario where two couplings are turned on simultaneously, again setting the remaining couplings to their SM values. We focus on a comparison of single-$\beta$ and $0\nu\beta\beta$ decay experiments. We show contours in the $v^2C_V^-$-$v^2C_A^-$ plane in the left panel of Fig.\ \ref{fig:2Dplots}, corresponding to a case in which right-handed neutrinos couple to quarks through (axial) vector currents. The single-$\beta$ bounds are depicted in blue and indicate good agreement with the SM expectation, which is represented by the origin. The blue contour resides within a square with boundaries $-0.03 < v^2C_{A,V}^- < 0.03$. In the case of Majorana neutrinos, the (future) $0\nu\beta\beta$ constraints are shown by the (dashed) red contours, we again assumed there are no cancelations between the $\sim m_{\bt\bt}$ and non-standard contributions, while we used the NMEs of Ref.\ \cite{Hyvarinen:2015bda} and neglected the $\Or(1)$ hadronic and nuclear uncertainties.
The bounds span an oval that can roughly be described by $\left(v^2 C_A^-\right)^2+2.2\left(v^2C_V^-\right)^2\lesssim 0.05\times \frac{120\,{\rm eV}}{m_4}$. The current $0\nu\bt\bt$ constraints are thus competitive with  single-$\bt$ probes for $m_4 > 200$ eV or so. Next-generation experiments will push this to $m_4 >20$ eV. 

The right panel of Fig.\ \ref{fig:2Dplots} instead shows the $v^2C_A^-$-$v^2C_T^-$ plane where the (not too significant) discrepancy found in Ref.~\cite{Falkowski:2020pma} is visible. If this anomaly would be confirmed, the complementary nature of $0\nu\beta\beta$ could prove useful in identifying the nature and mass of right-handed neutrinos. With the current experimental $0\nu\bt\bt$ limits a nonzero value of $C_{A,T}^-$, consistent with the $\beta$-decay data, require neutrinos to be either Dirac or have masses below $300$ eV. Next-generation $0\nu\beta\beta$ experiments will bring this down to $30$ eV. Complementary constraints on neutrino masses in the eV range come from oscillation experiments \cite{Kopp:2013vaa,Falkowski:2019xoe}, while searches for kinks in the electron spectrum of single-$\beta$ decay experiments \cite{Shrock:1980vy,Bolton:2019pcu} are sensitive to masses up to the MeV range. 

\begin{figure}[t!]\begin{center}
\includegraphics[scale=0.55]{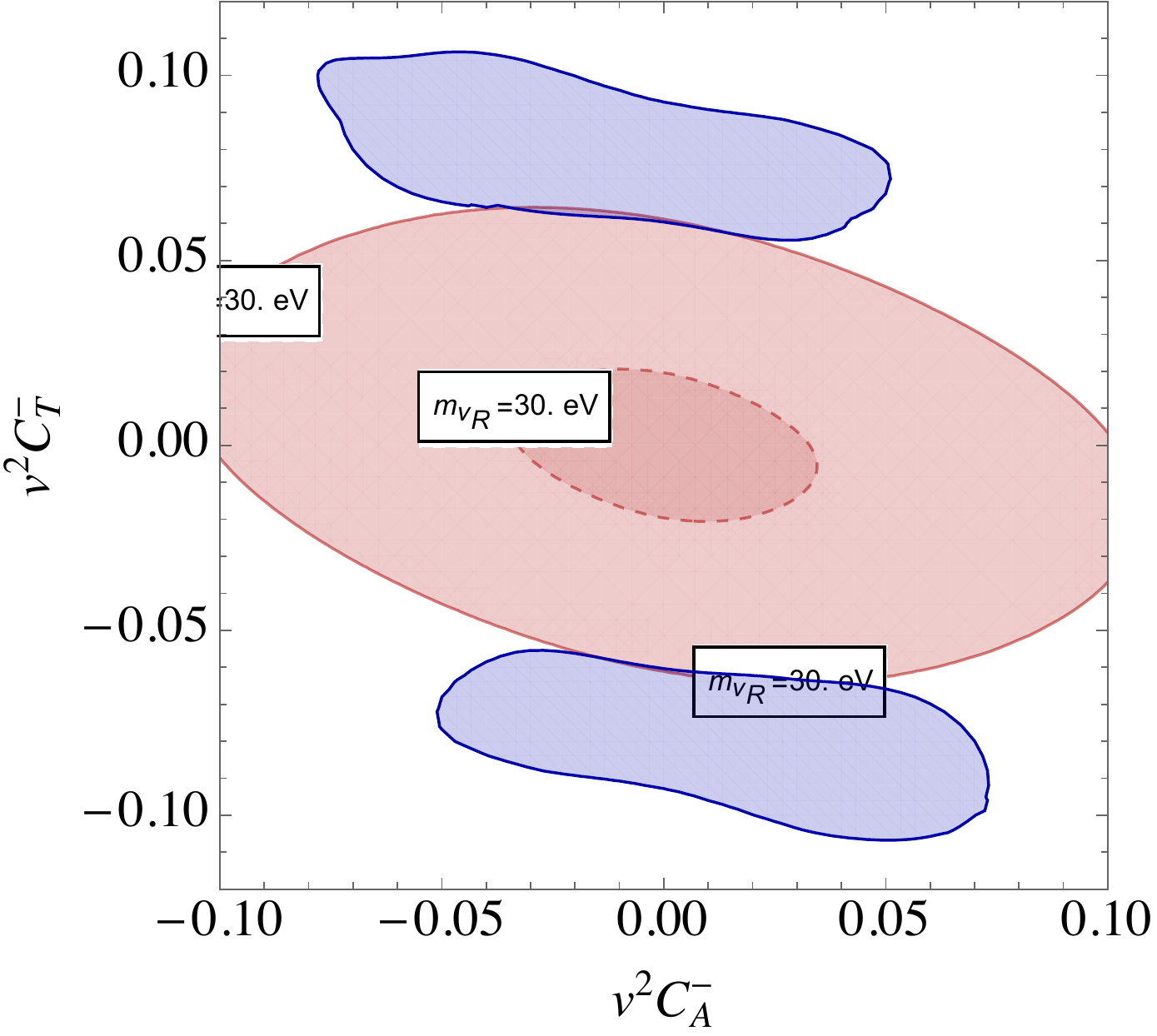}
\end{center}
     \caption{
Similar to Fig.\ \ref{fig:2Dplots}, however, here the single-$\bt$ contours have been obtained by marginalizing over the other Wilson coefficients.
}
       \label{fig:2Dmarg}
\end{figure}

Finally, we use the same machinery to perform a fit of the $\beta$-decay data, where the single-$\beta$ data is marginalized over all 5 couplings, while we only consider the $C_{A,T}^-$ contributions to $0\nu\bt\bt$. The resulting contour is shown in Fig.\ \ref{fig:2Dmarg}, which more clearly shows a preference for nonzero $C_T^-$. In this case, the preferred fit values of the Wilson coefficients are only consistent with Majorana neutrinos if the sterile neutrino has a mass below $30$ eV. Next-generation experiments will push this $3$ eV, covering essentially the entire possible mass range. 

\section{A leptoquark model}\label{LQ}
So far our analysis was performed purely in an EFT framework and we constrained effective couplings of the massive sterile neutrino. While this is convenient to demonstrate the complementarity between single-$\beta$ and $0\nu\beta\beta$ experiments, it might also lead to oversimplifications. 
To get a sense of more realistic BSM scenarios that induce multiple dimension-six operators at the same time, we consider a toy model with several leptoquarks (LQs) that have interactions with right-handed neutrinos. Integrating out the leptoquarks leads to dimension-six $\nu$SMEFT operators \cite{Bischer:2019ttk}. 

Following the notation of Ref.~\cite{Dorsner:2016wpm} we consider a BSM model with two scalar LQs, $S_1$ and $\tilde R_2$. $S_1$ has mass $M_1$ and transforms under $SU(3)_c$ as an anti-triplet, under $SU(2)_L$ as a singlet, and carries nonzero hypercharge$: S_1 \left({\bf \bar 3},~{\bf 1},~1/3\right)$, while $\tilde R_2$ has mass $M_2$ and transforms under $SU(3)_c$ as a triplet, under $SU(2)_L$ as a doublet, and also carries nonzero hypercharge$: \tilde R_2 \left({\bf 3},~{\bf 2},~1/6\right)$. The relevant couplings to fermions can be written as
\begin{align}
\label{eq:LQlag}
{\cal L}_{\rm LQ_s} &= \left[\bar{Q}_{L}^{c\,i} y^{LL} \epsilon^{ij}L_{L}^{j}
+ \bar{d}_{R}^{c} y^{\overline{RR}}\nu_{R}\right]S_{1} \nonumber\\[5pt]
&+\left[\bar{Q}^{i}_{L} y^{\overline{LR}}\nu_{R} - \bar{d}_{R}\epsilon^{ij} y^{RL}L_{L}^{j}\right]\tilde R^i+ \textrm{h.c.}\,,
\end{align}
where $y^{LL,RL}$ and $y^{\overline{RR},\overline{LR}}$ are $3\times 3$ and $3\times n$ Yukawa matrices, respectively, while  $i,j$ are $SU(2)_L$ indices. Integrating out the two LQs at tree level gives rise to dim-6 operators and at low energies the effective interactions between neutrinos and the first-generation quarks and charged-leptons become
\begin{align}
{\cal L }^{(6)} \supset \frac{2G_F}{\sqrt{2}}\bigg[ c_{\substack{{\rm VL}\\e\al}}^{(6)} (\bar u_L \gamma^\mu d_L) (\bar e_{L}  \gamma_\mu \nu_{L\, \al}) + \bar{c}_{\substack{{\rm SR}\\ea}}^{(6)} (\bar{u}_L d_R) (\bar{e}_L\nu_{R\,a}) + \bar{c}^{(6)}_{\substack{{\rm T}\\ea}} (\bar{u}_L \sigma^{\mu\nu}d_R) (\bar{e}_L\sigma^{\mu\nu}\nu_{R\,a}) \bigg]\,,
\end{align}
where we neglected neutral-current operators that are not relevant for single- and double-$\bt$ decays.
Furthermore, $\al=1,\dots 3$ and $a=1,\dots n$ indicate the active and sterile neutrinos and the Wilson coefficients are given by
\begin{align}
c_{\substack{{\rm VL}\\e\al}}^{(6)} &= -2V_{ud}\dt_{e\al} - \frac{v^2}{2 M_1^2}y_{1e}^{LL*} y_{1\al}^{LL} \,,\nnl[5pt]
\bar{c}_{\substack{{\rm SR}\\ea}}^{(6)} &= 
\frac{v^2}{2} \left( \frac{y^{\overline{LR}}_{1a} \, y^{RL*}_{1e}}{M_2^2} - \frac{y^{\overline{RR}}_{1a} \, y^{LL*}_{1e}}{M_1^2} \right) \,,\nnl[5pt]
\bar{c}^{(6)}_{\substack{{\rm T}\\ea}} &= \frac{v^2}{8} \left( \frac{y^{\overline{LR}}_{1a} \, y^{RL*}_{1e}}{M_2^2} + \frac{y^{\overline{RR}}_{1a} \, y^{LL*}_{1e}}{M_1^2} \right)\,.
\end{align}
In the mass basis the matching coefficients become
\begin{align}
\left(C^{(6)}_{\rm VLL}\right)_{ei} &= \sum_{\al=1}^3c_{\substack{{\rm VL}\\e\al}}^{(6)}  U_{\al i}\,, \nnl[5pt]
\left(C^{(6)}_{\rm SRR}\right)_{ei} &= \sum^n_{a=1} \bar{c}^{(6)}_{\substack{{\rm SR}\\ ea}}U^*_{3+a,i} \,,\nnl[5pt]
\left(C^{(6)}_{\rm TRR}\right)_{ei} &= \sum^n_{a=1} \bar{c}^{(6)}_{\substack{{\rm T}\\ ea}}U^*_{3+a,i}\,.
\end{align}

A  simplification is made possible by the fact that, as pointed out in Ref.\ \cite{Gonzalez-Alonso:2018omy}, the Wilson coefficient $C^{(6)}_{\rm VLL}$ cannot be independently determined by the nuclear data alone. In particularly, in Eq.\ \eqref{eq:TH_LYtoRWEFT}, it is {\em not} possible to distinguish the pure CKM element $V_{ud}$ from the new physics contamination parameterized by $V_{ud}\,(1 + \epsilon_L + \epsilon_R)$.
\footnote{The data only constrain four Wilson coefficients $C_{V,A,S,T}^+$, which depend on five parameters: $V_{ud}$ and $\epsilon_{L,R,S,T}$, leaving one flat direction.}
We therefore absorb $C^{(6)}_{\rm VLL}$ into the CKM element $V_{ud}$, and only consider two Wilson coefficients $C^{(6)}_{\rm SRR}$ and $C^{(6)}_{\rm TRR}$ in this simple LQ model.

 \begin{figure}[t!]\begin{center}
\includegraphics[scale=0.55]{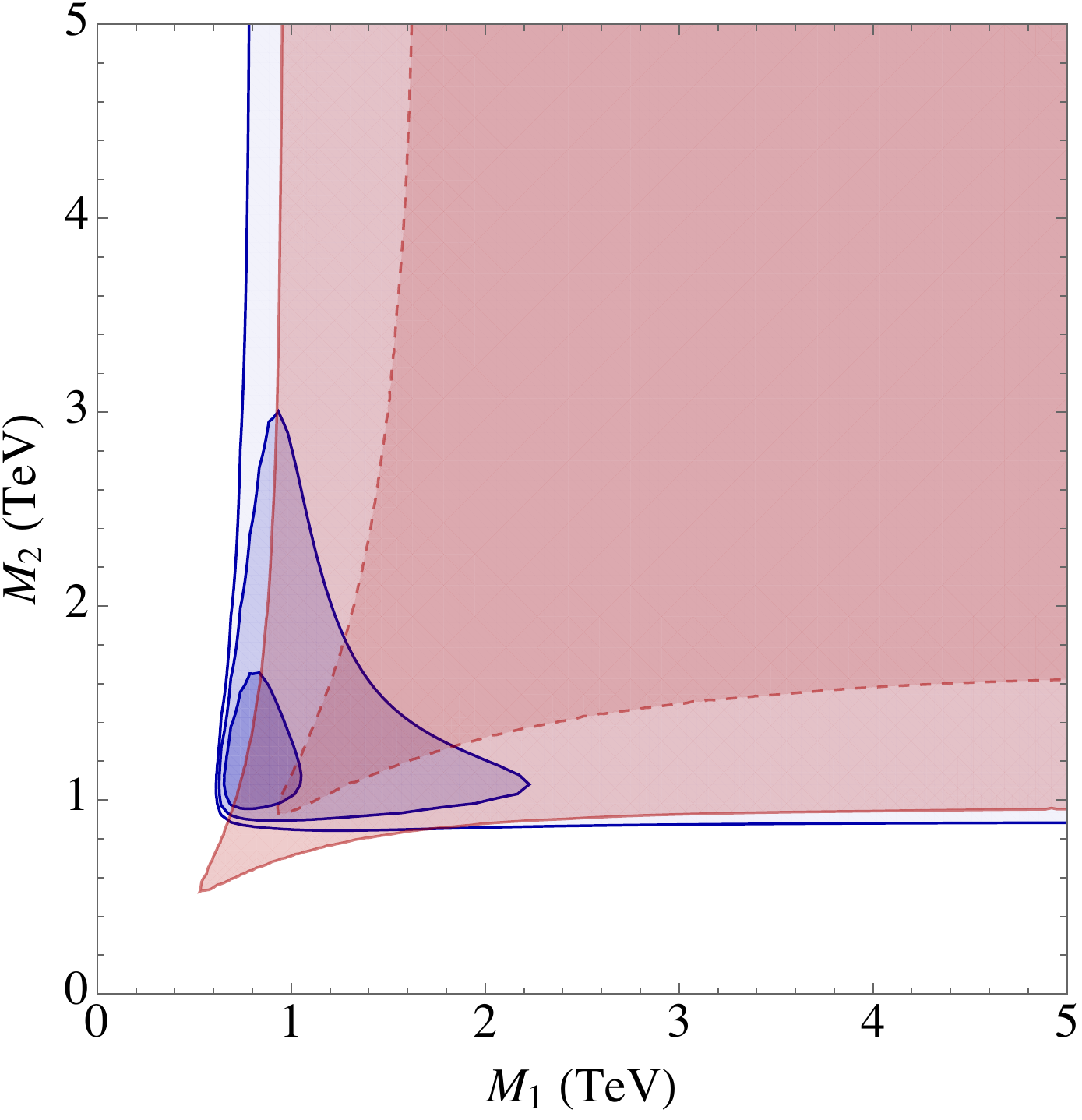}
\end{center}
     \caption{
The red (dashed) line shows the current (future) limit from $0\nu\bt\bt$ for $m_{\nu_R} = 10$ eV. The blue regions denote the $\Delta \chi^2=1,2,3$ contours from the single-$\bt$ decay fit.
}
       \label{fig:LQ}
\end{figure}

We show the resulting allowed regions in the $M_1-M_2$ plane in Fig.\ \ref{fig:LQ}, where we set $ y^{\overline{LR}}_{11} \, y^{RL*}_{1e}=y^{\overline{RR}}_{11} \, y^{LL*}_{1e}=1$ for simplicity. Here the red (dashed) contours show the (future) $0\nu\bt\bt$ limits for $m_4 = 10$ eV, while the blue regions denote the $\Delta \chi^2=1,2,3$ contours from single-$\bt$ decays.
These $0\nu\bt\bt$ limits were obtained assuming there are no significant cancelations between the $\sim m_{\bt\bt}$ and non-standard contributions.  
In addition, we used the NMEs from Ref.\ \cite{Hyvarinen:2015bda} and do not show the theoretical uncertainties due to the nuclear and hadronic matrix elements.
The figure clearly shows the preference of the single-$\bt$ fit for a nonzero tensor coupling, $C_T^-$, represented by the closed contours at $\Delta \chi^2=1,2$, while for $\Delta \chi^2=3$ the fit allows for $C_T^-\to0$ and for the LQ masses to decouple.
By assuming that neutrinos are Majorana one can again compare to the current $0\nu\bt\bt$ limits. From the figure, one can see that these constraints are already cutting into the region preferred by single-$\bt$ decay at $\Delta \chi^2=1$, for  $m_{\nu_R} = 10$ eV. The prospective EXO-200 measurements would allow one the completely exclude the $\Delta \chi^2=1$ region and large parts of the $\Delta \chi^2=2$ parameter space.

It should be mentioned that the considered scenario does not provide a complete description of neutrino masses or attempt to reconcile with direct collider observables which typically exclude LQs with masses below $1-2$ TeV \cite{CMS:2018ncu,ATLAS:2020dsk}. A realistic model would therefore require more involved model building beyond the relatively simple extension of the SM considered here. 
We nevertheless consider the above scenario as a toy model as it clearly shows the complementary information on the nature and size of neutrino masses $0\nu\bt\bt$  could provide once a signal of nonzero BSM couplings would be found in probes of single $\bt$.

\section{Conclusions}\label{conclusion}

Sterile neutrinos play a role in numerous promising extensions of the SM and are strongly motivated by the necessity to generate neutrino masses. If sterile neutrinos exist, it is very possible that neutrinos are Majorana particles as no SM symmetries forbid sterile Majorana masses. In broad classes of SM extensions, sterile neutrinos only appear sterile at low energies, but interact through the exchange of heavy beyond-the-Standard-Model particles. Such SM extensions include left-right symmetric models, GUTs, leptoquark, and $Z'$ models. In this work we studied how such interactions can be probed in $\beta$- and $0\nu\beta\beta$-decay experiments and in particular how these experiments complement each other in determining the mass and nature, Dirac or Majorana, of neutrinos. 

To do so we employed the analysis of Ref.~\cite{Falkowski:2020pma} which performed a comprehensive analysis of non-standard neutrino interactions in $\beta$-decay experiments. They demonstrated that precision $\beta$-decay experiments probe non-standard couplings to sterile neutrinos at the $\mathcal O(10^{-1}-10^{-2})$ level, corresponding to BSM scales of a few TeV. They also found a slight hint for a nonzero tensor coupling to sterile neutrinos. We addressed what these results imply for the search of sterile neutrinos with $0\nu\beta\beta$ decay experiments. On the $0\nu\bt\bt$ side, we calculated the decay rates as a function of sterile neutrino masses and non-standard couplings at the dimension-six level of the neutrino-extended Standard Model Effective Field Theory. We focussed on rather light sterile neutrinos as only these can be produced in neutron and nuclear $\beta$-decay processes.
It is worthwhile to mention that the KATRIN experiment, which measures the tritium $\beta$-decay spectrum, may also provide competitive constraints on the non-standard couplings of the sterile neutrino within our mass range. Given the current limits, the non-standard couplings could induce relative distortions of the spectrum at the per mille level \cite{Ludl:2016ane}. As was shown for a benchmark calculation with $m_{\nu_R}\sim 5$ keV, these effects could allow for improved constraints assuming that the whole electron energy spectrum of KATRIN is accessible and theoretical and systematical uncertainties can be reduced below the per mille level.

For sufficiently small masses, $\beta$-decay constraints are neutrino-mass independent whereas $0\nu\beta\beta$ decay rates scale quadratically with the Majorana mass. We showed that, barring cancelations between $\sim m_{\bt\bt}$ and non-standard contributions, $0\nu\bt\bt$ measurements start to become competitive for right-handed neutrino masses larger than about $100$ eV with a mild dependence on the particular interaction under consideration.
Under the same assumptions, we also discussed the range of neutrino masses for which one would expect a measurable signal in $0\nu\beta\beta$, assuming couplings of right-handed neutrinos were to be confirmed in either  single $\bt$ decays or at the LHC. For right-handed neutrino couplings of the size indicated by the hint in the recent $\bt$-decay results, we show that next-generation $0\nu\beta\beta$ experiments would be able to determine the Dirac or Majorana nature of neutrinos for $m_{\nu_R}\gtrsim 3$ eV. Finally, we discussed the implications of the hints in single-$\bt$ and the $0\nu\beta\beta$ limits in the context of a specific (toy) model of BSM physics involving two leptoquarks. Such a model with TeV leptoquark masses can account for the discrepancy and imply a measurable signal in next-generation $0\nu\beta\beta$ experiments unless neutrinos are Dirac particles. 

Our study demonstrates the complementary nature of different neutrino probes, and the various aspects of the theory space they individually are sensitive to. In case of a future discovery of a deviation from the Standard Model predictions in one of the experiments, looking at other probes in a global approach provides valuable information to better understand the nature of neutrinos.

\section*{Acknowledgments}
We thank Adam Falkowski and Mart\'in Gonz\'alez-Alonso for providing us with the $\chi^2$ function used in their work \cite{Falkowski:2020pma}, as well as Vincenzo Cirigliano and Emanuele Mereghetti for useful discussions and feedback on this work. 
WD is supported by  U.S.\ Department of Energy Office of Science, under contract DE-SC0009919.
TT is supported in part by  U.S.\ Department of Energy Office of Science, under contract DE-SC0011640.

\appendix
\section{$\nu$SMEFT operators and matching}\label{app:match}
Here we briefly discuss the dimension-six operators making up the $\nu$SMEFT Lagrangian in Eq.\ \eqref{eq:smeft} and give the matching of these operators onto the low-energy Lagrangian of Eq.\ \eqref{6final}. The dimension-six Lagrangian involving left- and right-handed neutrinos, $\mathcal L^{(\bar 6)}_{\nu_{L,R}}$, is given by the sum of the operators (and their hermitian conjugates) in Table \ref{tab:O6L} and Table \ref{tab:O6R}, respectively, where each operator is multiplied by a corresponding Wilson coefficient.
The matching contributions of these Wilson coefficients to the couplings to left-handed neutrinos can then be written as,
\bea\label{match6LNC}
C_{\rm VLL}^{(6)} &=& \Bigg[-2V_{ud}\mathbb{1}+2v^2\left[C_{LQ\,3}^{(6)}-C_{HL\,3}^{(6)}-C_{HQ\,3}^{(6)}\, \mathbb{1}\right]-\frac{4\sqrt{2}v}{g }M_e \left(C^{(6)}_{eW}\right)^\dagger\nn\\
&&-\frac{4\sqrt{2}v}{g} C^{(6)}_{\nu W} M_D^\dagger\Bigg]PU-\frac{4\sqrt{2}v}{g} C_{\nu W}^{(6)}M_R^\dagger \,P_sU\,,\nn\\
C_{\rm VRL}^{(6)} &=&-v^2C_{Hud}^{(6)}\, PU,\nn\\
C_{\rm SRL}^{(6)} &=& v^2 \left(C_{LedQ}^{(6)}\right)^\dagger \,PU\,,\nn\\
C_{\rm SLL}^{(6)} &=& v^2\left(C_{LeQu\,1}^{(6)}\right)^\dagger\, PU\,,\nn\\
C_{\rm TLL}^{(6)} &=& v^2\left(C_{LeQu\,3}^{(6)}\right)^\dagger \,PU\,,
\eea
where the first contribution to $C_{\rm VLL}^{(6)}$ is the SM contribution. 

The Wilson coefficient of the operators involving right-handed chiralities are,
\bea\label{match6LNCsterile}
 C_{\rm VLR}^{(6)} &=& \left[-v^2C_{H\nu e}^{(6)}- \frac{4\sqrt{2} v}{g}   \left(C_{\nu W}^{(6)}\right)^\dagger M_e-\frac{4\sqrt{2}v}{g}M_D^\dagger C_{eW}^{(6)} \right]^\dagger P_s U^*\,,\nn\\
 C_{\rm VRR}^{(6)} &=& v^2\left(C_{du\nu e}^{(6)}\right)^\dagger \,P_s U^*\,,\nn\\
 C_{\rm SRR}^{(6)}&=& \left[-v^2C_{L\nu Qd}^{(6)}+\frac{v^2}{2} C_{LdQ\nu }^{(6)}\right]P_s U^*\,,\nn\\
 C_{\rm SLR}^{(6)}&=& v^2\left(C_{Qu\nu L}^{(6)}\right)^\dagger \,P_s U^*\,,\nn\\
 C_{\rm TRR}^{(6)} &=& \frac{v^2}{8} C_{LdQ\nu }^{(6)}\,P_s U^*\,.
\eea

{\renewcommand{\arraystretch}{1.3}\begin{table}[t!]\small
\center
\begin{tabular}{||c|c||c|c||c|c||}
\hline Class $1$& $\psi^2 H X$ & Class $2$ & $\psi^2 H^2 D$        & Class $3$ &  $\psi^4 $\\
\hline
$\mathcal O_{eW}^{(6)}$  &        $(\bar L  \sigma^{\mu\nu} e) \tau^I H W_{\mu\nu}^I$ & 
 $ \mathcal O_{H L\,3}^{(6)}$ &   $(H^\dag i\overleftrightarrow{D}^I_\mu H)(\bar L \tau^I \gamma^\mu L)$   &
$\mathcal O_{LeQu\,1}^{(6)}$&  $(\bar L ^j e) \epsilon_{jk} (\bar Q^k u)$  \\
$\mathcal O_{uW}^{(6)}$  &        $(\bar Q \sigma^{\mu\nu} u) \tau^I \widetilde H \, W_{\mu\nu}^I$ &
$\mathcal O_{H Q\,3}^{(6)}$      & $(H^\dag i\overleftrightarrow{D}^I_\mu H)(\bar Q \tau^I \gamma^\mu Q)$ &
$\mathcal O_{LeQu\,3}^{(6)}$ &    $(\bar L^j \sigma_{\mu\nu} e) \epsilon_{jk} (\bar Q^k \sigma^{\mu\nu} u)$\\
$\mathcal O_{dW}^{(6)}$ & $(\bar Q \sigma^{\mu\nu} d) \tau^I H\, W_{\mu\nu}^I$ &
$\mathcal O_{H u d}^{(6)}$   & $i(\widetilde H ^\dag D_\mu H)(\bar u \gamma^\mu d)$ &
$\mathcal O_{LQ\,3}^{(6)}$     & $(\bar L \gamma^\mu \tau^I L)(\bar Q \gamma_\mu \tau^I Q)$ \\
 &  & & & $\mathcal O_{LedQ}^{(6)}$ & $(\bar L^j e)(\bar d Q^{j})$ \\\hline
\hline 
\end{tabular}
\caption{LNC {dim-6} operators \cite{Grzadkowski:2010es} involving active neutrinos that affect \NLDBD\ at tree level.}   \label{tab:O6L}
\end{table}}

{\renewcommand{\arraystretch}{1.3}\begin{table}[t!]\small
\center
\begin{tabular}{||c|c||c|c||}
\hline Class $1$& $\psi^2 H^3$  & Class $4$ &  $\psi^4 $\\
\hline
$\mathcal{O}^{(6)}_{L\nu H}$ & $(\bar{L}\nu_R)\tilde{H}(H^\dagger H)$ & $\mathcal{O}^{(6)}_{du\nu e}$ & $ (\bar{d}\gamma^\mu u)(\bar{\nu_R }\gamma_\mu e)$  \\ \cline{1-2}
 Class $2$&  $\psi^2 H^2 D$ &  $\mathcal{O}^{(6)}_{Qu\nu L}$ & $(\bar{Q}u)(\bar{\nu}_RL)$  \\ \cline{1-2}
$\mathcal{O}^{(6)}_{H\nu e}$ & $(\bar{\nu }_R\gamma^\mu e)({\tilde{H}}^\dagger i D_\mu H)$ & $\mathcal{O}^{(6)}_{L\nu Qd}$ & $(\bar{L}\nu_R )\epsilon(\bar{Q}d))$ \\ \cline{1-2}
Class $3$ & $\psi^2 H^3 D$  & $\mathcal{O}^{(6)}_{LdQ\nu }$ & $(\bar{L}d)\epsilon(\bar{Q}\nu_R )$ \\ \cline{1-2}
$\mathcal{O}^{(6)}_{\nu W}$ &$(\bar{L}\sigma_{\mu\nu}\nu_R )\tau^I\tilde{H}W^{I\mu\nu}$  & &\\
\hline
\end{tabular}
\caption{LNC {dim-6} operators \cite{Liao:2016qyd} involving a sterile neutrino that affect \NLDBD\ at tree level.
} \label{tab:O6R}
\end{table}}

\section{Details of the $0\nu\bt\bt$ calculation}\label{app:double}
Here we briefly discuss the ingredients involved in the calculation of $0\nu\bt\bt$ decay rates that were not fully explained in Sect.\ \ref{0vbb}. We start with the definition of the nuclear matrix elements (NMEs), after which we discuss the subamplitudes, $\mathcal A_{L,R,M}^{(\nu)}(m_i)$, that are induced by the exchange of hard neutrinos.

\subsection{Nuclear matrix elements}
The NMEs used in Sect.\ \ref{0vbb} are defined as follows,
\begin{eqnarray}\label{m1}
\mathcal M_{V}(m_i) &=&  -\frac{g_V^2}{g_A^2} M_F(m_i) + M^{MM}_{GT}(m_i) + M^{MM}_{T}(m_i)\,, \nn \\ 
\mathcal M_{A}(m_i) &=&  M^{AA}_{GT}(m_i) + M^{AP}_{GT}(m_i)  +M^{PP}_{GT}(m_i)  +  M^{AP}_{T}(m_i) + M^{PP}_{T}(m_i)\,, \nn\\
\mathcal M_{PS}(m_i)     &=& \frac{1}{2} M^{AP}_{GT}(m_i) + M^{PP}_{GT}(m_i)  + \frac{1}{2} M^{AP}_{T}(m_i) + M^{PP}_{T}(m_i)\, ,\nn \\
\mathcal M_{PS, sd} &=& \frac{1}{2} M^{AP}_{GT,\,sd}(0) + M^{PP}_{GT,\, sd}(0)  + \frac{1}{2} M^{AP}_{T,\,sd}(0) + M^{PP}_{T,\, sd}(0)\, ,\nn \\
\mathcal M_{S}(m_i) &=& \frac{g^2_S}{g^2_A} M_F(m_i) 
\, ,\nn \\
\mathcal M_{T}(m_i) &=& 16\frac{g^2_T}{g^2_A} M^{AA}_{GT}(m_i) \, , \nn\\
 \mathcal M_{T\, V}(m_i) &=& -4\frac{g^\prime_T  g_V
  }{g_A^2} \frac{m^2_\pi}{m_N^2} M_{F,\, sd}(m_i)  + \frac{16 g_T}{ g_M} \left[ M_{GT}^{MM}(m_i) + M_T^{MM}(m_i)\right]\,, \nn \\
\mathcal M_{V\,A}(m_i) &=& 2\frac{g_A}{g_M}  \left[M_{GT}^{MM}(m_i) + M_{T}^{MM}(m_i) \right]\,, \nn \\
\mathcal M_{T\,A}(m_i) &=& \frac{g_T}{g_A}  \left[2M_{GT}^{AA}(m_i) + M_{GT}^{AP}(m_i) + M_T^{AP}(m_i) \right]\,,
\end{eqnarray}
in terms of a set of NMEs and the low-energy constants $g_T = 0.99 \pm 0.03$, $g_M = 4.7$, and the unknown $g_T' = \mathcal O(1)$ which we set to $1$ in this work.
The small $m_i$ limits of the NMEs have been calculated in the literature. We show the numerical values of $M_K^a(0)$ and $M_{K,\, sd}^a(0)$, with $K = F,GT,T$ and $a = AA,AP,PP,MM$, for several calculations and different isotopes in Table \ref{tab:comparison}. 

To obtain the $m_i$ dependence of these matrix elements we follow Refs.\ \cite{Dekens:2020ttz,Barea:2015zfa} and interpolate between the low- and high-mass regimes, where the relevant matrix elements are available. This is possible since $M_K^a(0)$ have been determined in the literature, while the large-mass limits are in most cases are related to the short-distance matrix elements, $M_{K,\, sd}^a(0)$.
For example, for the Fermi matrix element we use
\be\label{intsimple}
M_{F\,\mathrm{int}}(m_i) = M_{F,\, sd}\frac{m_\pi^2}{m_i^2 + m_\pi^2 \frac{M_{F,\, sd}}{M_F}}\,,
\ee
which has the correct $m_i\to0$ and $m_i\to \infty$ behavior.
We construct analogous expressions for $M^{AA}_{GT,T\,\mathrm{int}}(m_i)$, $M^{AP}_{GT,T\,\mathrm{int}}(m_i)$, and $M^{PP}_{GT,T\,\mathrm{int}}(m_i)$.  The $m_i$ dependence of the remaining matrix elements in Table~\ref{tab:comparison} can be related to the previously discussed NMEs.
For the magnetic GT matrix element we have,
\bea
M^{MM}_{GT}(m_i) &=& \frac{g_M^2}{6 g_A^2} \left[\frac{m_\pi^2}{m_N^2} M^{AA}_{GT,\,sd}  - \frac{m_i^2}{m_N^2}M^{AA}_{GT}(m_i)\right]\,,
\eea
and the for short-distance NMEs
\bea
M_{F,\,sd}(m_i) &=& M_{F,\,sd}- \frac{m_i^2}{m_\pi^2}M_{F}(m_i)\,,\nn\\
M^{ab}_{GT,\,sd}(m_i) &=& M^{ab}_{GT,\,sd}- \frac{m_i^2}{m_\pi^2}M^{ab}_{GT}(m_i)\,,\nn\\
M^{ab}_{T,\,sd}(m_i) &=& M^{ab}_{T,\,sd}- \frac{m_i^2}{m_\pi^2}M^{ab}_{T}(m_i)\,.
\eea

\begin{table}
\center
$\renewcommand{\arraystretch}{1.5}
\begin{array}{l||rrr|rr|rr |rr}
 \text{NMEs} & \multicolumn{3}{c|}{\text{}^{76} \text{Ge}} & \multicolumn{2}{c|}{\text{}^{82} \text{Se}} & \multicolumn{2}{c|}{ \text{}^{130} \text{Te}} & \multicolumn{2}{c}{  \text{}^{136} \text{Xe}}  \\
& \text{\cite{Hyvarinen:2015bda}} &   \text{\cite{Menendez:2017fdf}}  
& \text{\cite{Barea:2015kwa,Barea}} & \text{\cite{Hyvarinen:2015bda}} &    \text{\cite{Menendez:2017fdf}} & \text{\cite{Hyvarinen:2015bda}} &   \text{\cite{Menendez:2017fdf}} & \text{\cite{Hyvarinen:2015bda}} &   \text{\cite{Menendez:2017fdf}} \\
 \hline
 M_F 			   & $-$1.74    &  $-$0.59 	& $-$0.68 	& $-$1.29 	&  $-$0.55	& $-$1.52    	&   $-$0.67	& $-$0.89  	&   $-$0.54 \\
 M_{GT}^{AA} 		   & 5.48       &  3.15	    	& 5.06 		& 3.87 		&  2.97		& 4.28    	&   2.97	& 3.16   	&   2.45 \\
 M_{GT}^{AP} 		   & $-$2.02    & $-$0.94	& $-$0.92 	& $-$1.46    	& $-$0.89   	& $-$1.74 	&  $-$0.97	& $-$1.19   	&  $-$0.79  \\
 M_{GT}^{PP} 		   & 0.66  	&  0.30		& 0.24 		& 0.48          &  0.28 	& 0.59   	&   0.31   	& 0.39   	&   0.25\\
 M_{GT}^{MM} 		   & 0.51 	& 0.22		& 0.17 		& 0.37       	& 0.20 		& 0.45 		&  0.23 	& 0.31 		&  0.19 \\
 M_T^{AA} 	   	   &  -    	& - 		& - 		&  -         	& -		&  -     	&  -      	&  -     	&   -	\\
 M_T^{AP} 	   	   & $-$0.35 	& $-$0.01	& $-$0.31 	& $-$0.27    	& $-$0.01 	& $-$0.50	&     0.01	& $-$0.28 	&     0.01	\\
 M_T^{PP} 	 	   & 0.10     	&    0.00	& 0.09  	& 0.08       	& 0.00 		& 0.16 		&  $-$0.01 	& 0.09 		&  $-$0.01 	 \\
 M_T^{MM}           	   & $-$0.04	&   0.00	& $-$0.04	& $-$0.03    	&   0.00 	& $-$0.06 	&    0.00	& $-$0.03 	&    0.00  \\\hline
 M_{F,\, sd}  	   	   & $-$3.46 	& $-$1.46	& $-$1.1  	& $-$2.53   	& $-$1.37 	& $-$2.97 	&  $-$1.61 	& $-$1.53   	&  $-$1.28	\\
 M^{AA}_{GT,\, sd}  	   &    11.1 	& 4.87		& 3.62		& 7.98    	& 4.54 		& 	10.1 	&  5.31 	&    5.71    	&  4.25  \\
M^{AP}_{GT,\, sd}	   & $-$5.35 	& $-$2.26 	& $-$1.37 	& $-$3.82    	& $-$2.09 	& $-$4.94 	&  $-$2.51 	& $-$2.80  	&  $-$1.99  \\
M^{PP}_{GT,\, sd}	   & 1.99 	& 0.82		& 0.42		& 1.42      	& 0.77 		& 1.86 		&  0.92 	& 1.06  	&   0.74\\
M^{AP}_{T,\, sd}	   & $-$0.85 	&  $-$0.05	&  $-$0.97	& $-$0.65    	&  $-$0.05	& $-$1.50 	&   0.07	& $-$0.92  	&   0.05		\\  
M^{PP}_{T,\, sd} 	   & 0.32 	&  0.02		&  0.38		& 0.24       	&  0.02		& 0.58 		&   $-$0.02	& 0.36  	&   $-$0.02 \\
\end{array}$
\caption{
Comparison of NMEs computed in the quasi-particle random phase approximation  \cite{Hyvarinen:2015bda}, shell model  \cite{Menendez:2017fdf}, and interacting boson model  \cite{Barea:2015kwa,Barea}
for several isotopes  of experimental interest. See e.g.\ \cite{Deppisch:2020ztt} for a more recent calculation in the interacting boson model which, however, uses slightly differently defined matrix elements.}
\label{tab:comparison}
\end{table}

\subsection{Hard neutrino exchange}\label{app:hardnu}
The contributions due to hard neutrinos can be written as,
\bea\label{eq:harpipi}
\mathcal A_L^{(\nu)}(m_i) &=& \frac{m_\pi^2}{m_e v} \Bigg[\left(\frac{C_{i\,L}^{\pi\pi}}{m_\pi^2}+ C_{i\,L}^{\prime\pi\pi}\right) \mathcal M_{PS,sd} + \frac{C_{i\,L}^{\pi N}-C_{i\,L}^{\prime\pi\pi}}{2}  \left( M^{AP}_{GT, \, sd}+M^{AP}_{T, \, sd} \right) \nn\\
&&- \frac{2}{g_A^2}  C^{NN}_{i\,L} M_{F,\, sd} \Bigg]\,,\nn\\
\mathcal A_M^{(\nu)} &=& \frac{m_\pi^2}{m_e v}  \bigg[  - \frac{2}{g_A^2}C^{NN}_{i\,V}   \, M_{F,\, sd} 
+ \frac{1}{2}C_{i\,V}^{\pi N}
\left( M^{AP}_{GT, \, sd}  + M^{AP}_{T, \, sd} \right) \bigg]\,.\label{eq:ALhardNu}
\eea
in terms of the effective $\pi\pi$, $\pi N$, and $NN$ interactions, $C_{i}^{\pi\pi}$, $C_{i}^{\pi N}$, and $C_{i}^{NN}$. $\mathcal A_R^{(\nu)}(m_i)$ can be obtained by replacing $C_{i\, L}^{\al}\to C_{i\, R}^{\al}$.
In the case of the dimension-six couplings under consideration here, together with the approximation in which we drop the interference terms with the SM weak current as discussed in Sect.\ \ref{0vbb}, several of the above terms vanish. It turns out that the derivative pion terms as well as the pion-nucleon couplings go to zero with the above approximations, so that we have $C_{i\,L}^{\prime\pi\pi} = C_{i\,L}^{\pi N} = C_{i\,V}^{\pi N}=0$.
The remaining $C_\al^{\pi\pi}$  couplings are given by,
\bea
C_{i\, L,R}^{\pi\pi}&=&\frac{m_i v}{F_\pi^2}c_{i\,L,R}^{\nu\pi\pi}\,,\nn\\
c_{i\, L}^{\nu\pi\pi} &=&2g_{\rm LR}^{\rm \pi\pi}(m_i)\left(C_{\rm VLL}^{(6)}\right)_{ei}\left(C_{\rm VRL}^{(6)}\right)_{ei}-2g_{\rm S1}^{\pi\pi}(m_i)\left[\left(C_{\rm SLR}^{(6)}\right)^2_{ei}+\left(C_{\rm SRR}^{(6)}\right)^2_{ei}\right]\nn\\
&&+4g_{\rm S2}^{\pi\pi}(m_i)\left(C_{\rm SLR}^{(6)}\right)_{ei}\left(C_{\rm SRR}^{(6)}\right)_{ei}-2g_{\rm TT}^{\pi\pi}(m_i)\left(C_{\rm TRR}^{(6)}\right)^2_{ei}\,.
\eea
Here, all LECs scale as $g^{\pi\pi}_i=\Or(F_\pi^2)$ and the right-handed coupling $c_{i\, R}^{\nu\pi\pi} $ can be obtained from $c_{i\, L}^{\nu\pi\pi}$  by interchanging the $L,R$ labels on the Wilson coefficients, $L\leftrightarrow R$, while leaving those on the LECs unchanged.

The relevant $NN$ couplings can be written as,
\bea\label{eq:harNN}
C_{i\, L,R,V}^{ \rm N N}&=&\frac{m_i }{\Lambda_\chi}c_{i\,L,R,V}^{\nu {\rm N N}}\,,\nn\\
\frac{c_{i\, L}^{\nu NN}}{v\Lambda_\chi}&=&
\frac{g_{\nu}^{\rm NN}(m_i)}{4}\left[\left(C_{\rm VLL}^{(6)}\right)_{ei}^2+\left(C_{\rm VRL}^{(6)}\right)_{ei}^2\right]
+\frac{g_{\rm LR}^{\rm NN}(m_i)}{2}\left(C_{\rm VLL}^{(6)}\right)_{ei}\left(C_{\rm VRL}^{(6)}\right)_{ei}\nn\\
&&+\frac{g_{\rm S1}^{\rm NN}(m_i)}{4}\left[\left(C_{\rm SRR}^{(6)}\right)^2_{ei}+\left(C_{\rm SLR}^{(6)}\right)^2_{ei}\right]-\frac{g_{\rm S2}^{\rm NN}(m_i)}{2}\left(C_{\rm SRR}^{(6)}\right)_{ei}\left(C_{\rm SLR}^{(6)}\right)_{ei}\nn\\
&&+\frac{g_{\rm TT}^{\rm NN}(m_i)}{4}\left(C_{\rm TRR}^{(6)}\right)^2_{ei}\,,\nn\\
\frac{c_{i\, V}^{\nu {\rm NN}}}{v\Lambda_\chi}&=&g_{\rm SLL,VLL}^{\rm NN}(m_i)\left[\left(C_{\rm SLL}^{(6)}+C_{\rm SRL}^{(6)}\right)_{ei}\left(C_{\rm VLL}^{(6)}+C_{\rm VRL}^{(6)}\right)_{ei}-\left(L\leftrightarrow R \right)\right]\nn\\
&&+g_{\rm TLL,VLL}^{\rm NN}(m_i)\left[\left(C_{\rm TLL}^{(6)}\right)_{ei}\left(C_{\rm VLL}^{(6)}-C_{\rm VRL}^{(6)}\right)_{ei}-\left(L\leftrightarrow R\right)\right]\,.
\eea
The right-handed couplings $c_{i\, R}^{\rm NN, \nu{\rm NN}} $ can again be obtained from $c_{i\, L}^{\rm NN, \nu{\rm NN}} $  by interchanging the $L,R$ labels on the Wilson coefficients, $L\leftrightarrow R$, while leaving those on the LECs unchanged.  Note that some of the terms in Eqs.\ \eqref{eq:harpipi} and \eqref{eq:harNN} can be neglected given our assumptions, but can play a role in the right-handed couplings after the above replacement.
Using naive-dimensional analysis would lead one to suspect that the $g_\al^{NN}$ scale as $1/\Lambda_\chi^2$, where $\Lambda_\chi\sim1$ GeV is the breakdown scale of chiral perturbation theory. However, it turns out that these LECs need to be enhanced by $\Lambda_\chi^2/F_\pi^2$ in order to obtain renormalized amplitudes, so that $g^{NN}_\al\sim 1/F_\pi^2$, see Ref.\ \cite{Dekens:2020ttz} for more details.

Similar to the $m_i$ dependence of the NMEs, we use the low- and high-mass limits of the LECs to construct interpolation formulae for these hadronic matrix elements. The interpolation we use is of the form 
\bea\label{eq:intLECs}
 g_\al (m_i)\big|_{\rm naive} =  \frac{g_\al (0)}{1+g_\al (0)\left[\frac{m_0^2}{m_i^2}\bar g_\al (m_0)\right]^{-1}}\,,
\eea
where $m_0\simeq 2$ GeV is a scale at which the procedure of integrating out the heavy neutrino becomes reliable. The $\bar g_\al (m)$ are effective LECs, scaling as $1/m^2$, which are needed in the $m_i\geq m_0$ region to ensure the amplitude reproduces the result from integrating out the neutrino at the quark level. 
The above expression reduces to $g_\al(0)$ for $m_i\ll m_\pi$ and  $\frac{m_0^2}{m_i^2}\bar g_\al (m_0)$  for $m_i\to \infty$. The required input to these interpolation formulae are thus $g_\al(0)$ and  $\bar g_\al (m_i)$, where the latter depends on the matrix elements of the dimension-nine operators, $\sim (\bar ud)^2 \bar e e^c$, that arise after integrating out neutrinos at the quark level. For simplicity we neglect the QCD evolution of these matrix elements, which can in principle be captured by Eq.\ \eqref{eq:intLECs}.

Performing the matching to the amplitudes that result from integrating out neutrinos with $m_i\gtrsim 2$ GeV gives rise to the following expressions for the $\pi\pi$ LECs
\bea\label{naiveinterpipi}
g_{\rm LR}^{\pi\pi}(m_i) &=& g_{\rm LR}^{\pi\pi}(0) \frac{1}{1 -4 \frac{m_i^2}{F_\pi^2} g_{\rm LR}^{\pi\pi}(0) \left[g_4^{\pi\pi}(m_0)  \right]^{-1}  }\,,\nn \\
g_{\rm S1}^{\pi\pi}(m_i) &=& g_{\rm S1}^{\pi\pi}(0) \frac{1}{1 +  8 \frac{m_i^2}{F_\pi^2} g_{\rm S1}^{\pi\pi}(0) \left[g_2^{\pi\pi}(m_0)  - B^2\right]^{-1}  }\,,\nn \\
g_{\rm S2}^{\pi\pi}(m_i) &=& g_{\rm S2}^{\pi\pi}(0) \frac{1}{1 - 8 \frac{m_i^2}{F_\pi^2} g_{\rm S2}^{\pi\pi}(0) \left[g_5^{\pi\pi}(m_0)/2  - B^2\right]^{-1}  }\,,\nn \\
g_{\rm TT}^{\pi\pi}(m_i) &=& g_{\rm TT}^{\pi\pi}(0) \frac{1}{1 +  \frac{m_i^2}{F_\pi^2} g_{\rm TT}^{\pi\pi}(0) \left[4 g_3^{\pi\pi}(m_0)  + 2 g_2^{\pi\pi}(m_0)\right]^{-1}}\,.
\eea
Similar expressions can in principle be derived for the $NN$ LECs \cite{Dekens:2020ttz}, currently, however, none of the needed LECs are known in the $NN$ sector. The techniques of Refs.~\cite{Cirigliano:2020dmx,Cirigliano:2021qko} where the short-distance LECs for the $LL$ case were calculated can in principle be applied to non-standard interactions as well, but this has not been done so far. In our numerical analysis we therefore estimate the hard-neutrino contributions by considering the $\pi\pi$ terms, while neglecting the $NN$ pieces. We take the following values for the needed LECs,
\bea
g_{\rm LR}^{\pi\pi}(0)&=&g_{\rm TT}^{\pi\pi}(0)=-g_{\rm S1}^{\pi\pi}(0)  = -g_{\rm S2}^{\pi\pi}(0) = F_\pi^2\,,
\qquad g_{\al}^{NN}(0) = 0\,,\nn\\
g_2^{\pi\pi} &=& 2.0\, {\rm GeV}^2\,,\qquad g_3^{\pi\pi} = -0.62\, {\rm GeV}^2\,,\nn\\
g_4^{\pi\pi} &=& -1.9\, {\rm GeV}^2\,,\qquad g_5^{\pi\pi} = -8.0\, {\rm GeV}^2\,.
\eea
Here the values for $g_{\al}^{\pi\pi}(0)$ are consistent with NDA estimates, while we use the lattice QCD results of Ref.\ \cite{Nicholson:2018mwc} for $g_{2,3,4,5}^{\pi\pi}$ evaluated at $\mu= 2$ GeV.
\bibliographystyle{utphysmod}
\bibliography{bibliography}

\end{document}